\documentclass[a4paper,11pt]{article}
\usepackage{jheppub} 
\usepackage{lineno}
\usepackage{mathrsfs}
\usepackage{amsmath}
\usepackage{comment}

\usepackage[normalem]{ulem}
\usepackage[english]{babel}
\usepackage[autostyle]{csquotes}

\usepackage{tikz}
\usepackage{amsmath} 
\usepackage{mathrsfs} 
\usepackage{xfp} 
\usepackage[outline]{contour} 
\usetikzlibrary{decorations.markings,decorations.pathmorphing}
\usetikzlibrary{angles,quotes} 
\usetikzlibrary{arrows.meta} 
\contourlength{1.4pt}
\newcommand{\zb}{\bar{z}}
\newcommand{\wb}{\bar{w}}
\tikzset{>=latex} 
\tikzstyle{cone}=[black,line width=0.2,top color=blue!60!black!30,
bottom color=blue!60!black!50!red!30,shading angle=60,fill opacity=0.9]
\tikzstyle{cone back}=[black,line width=0.1,dash pattern=on 1pt off 1pt]
\tikzstyle{world line}=[red!80,line width=0.6]
\tikzstyle{world line t}=[blue!80,line width=0.6]
\tikzstyle{particle}=[black,line width=0.5]
\tikzstyle{photon}=[-{Latex[length=4,width=3]},black,line width=0.4,decorate,
decoration={snake,amplitude=0.9,segment length=4,post length=3.8}]
\tikzstyle{singularity}=[black,line width=0.6,decorate,
decoration={zigzag,amplitude=2,segment length=6.17}]
\tikzset{declare function={%
    penrose(\x,\c)  = {\fpeval{2/pi*atan( (sqrt((1+tan(\x)^2)^2+4*\c*\c*tan(\x)^2)-1-tan(\x)^2) /(2*\c*tan(\x)^2) )}};%
    penroseu(\x,\t) = {\fpeval{atan(\x+\t)/pi+atan(\x-\t)/pi}};%
    penrosev(\x,\t) = {\fpeval{atan(\x+\t)/pi-atan(\x-\t)/pi}};%
    kruskal(\x,\c)  = {\fpeval{asin( \c*sin(2*\x) )*2/pi}};
}}

\title{\boldmath A stress tensor for asymptotically flat spacetime
}
\author{Jay Bhambure$\,^*$}
\author{and Hare Krishna$\,^{**,}$${}^\dagger$}

\affiliation{* Center for Nuclear Theory,\\ ${}\ $ ${}\ $  SUNY, Stony Brook, NY 11794, USA\\
** C.N. Yang Institute for Theoretical Physics,\\ ${}\ ${}\ $ $ SUNY, Stony Brook, NY 11794, USA\\
${}\,\dagger$ Weinberg \, Institute,\\
${}\ $ ${}\ $  The University of Texas at Austin, TX 78712, USA}

\emailAdd{jay.bhambure@stonybrook.edu}\emailAdd{hkrishna.phy@gmail.com}

\abstract{In this article, we propose a procedure for calculating the boundary stress tensor of a gravitational theory in asymptotically flat spacetime. As a case study, the stress tensor correctly reproduces the Brown-York charges for the Kerr black hole i.e. mass and angular momentum. In asymptotically flat spacetime, there are asymptotic symmetries called BMS symmetries. We compute charges associated with these symmetries with our proposed stress tensor. These charges are compared with the ones obtained using the Wald-Zoupas prescription. Our result for the stress tensor can be interpreted as the expectation value for the boundary stress tensor of some appropriate theory.}
\begin{document}
\hfill       YITP-SB-2024-33

\maketitle

\flushbottom
\section{Introduction and summary}
Gravity in any spacetime is believed to be holographic. The Hamiltonian of gravity is a boundary term (on-shell). One can even say that holography is implicit in canonical gravity \cite{Raju:2019qjq}. This fact about canonical gravity has immensely enriched our understanding of gravity in asymptotically Anti-de Sitter (AdS) spacetime \cite{Witten:1998qj,Aharony:1999ti,Maldacena:1997re}. The dictionary has been established between the bulk fields living in AdS and the corresponding boundary operators. In particular, the gravitons in AdS spacetime are dual to the stress tensor of the boundary CFT. Hence, the stress tensor of the boundary CFT can be obtained using the holographic method. In asymptotically AdS spacetime, Balasubramanian and Kraus \cite{Balasubramanian:1999re} used the variational principle with suitable counterterms to find such a stress tensor. Then they used it to compute the conserved charges (Brown-York) of the isometries of spacetime \cite{Brown:1992br}.
Furthermore, CFT in even dimensions often has conformal and Weyl anomalies. These anomalies have been efficiently computed using holographic methods (see \cite{Bianchi:2001kw,Skenderis:2002wp,Henningson:1998gx,deHaro:2000vlm} for some of the early papers). In particular, it has been shown that the central charge of a $2d$ CFT is proportional to the $3d$ AdS length scale $l_{AdS}$ in Planck's units \cite{Brown:1986nw}.
\begin{eqnarray}
    C= \frac{3 \, l_{AdS}}{2 G_N}
\end{eqnarray}
This relation has been established by computing the transformed stress tensor for AdS$_3$ spacetime upon the action of asymptotic symmetries and comparing it with the transformation rule for the boundary stress tensor. The asymptotic symmetries of AdS$_3$ spacetime lead to the Virasoro algebra \footnote{The BMS$_3$ algebra can be constructed by contracting the Virasoro algebra.} of the $2d$ CFT. In AdS, these holographic methods have led to numerous developments in understanding CFTs in general. The monotonic c- and a-theorems were first established using holographic methods \cite{Myers:2010tj,Myers:2010xs}, and then the a-theorem was proven in the QFT framework by Komargodski et al \cite{Komargodski:2011vj}. In some recent work by Karateev et al. \cite{Karateev:2023mrb} $\Delta c-\Delta a$ is bound along the RG flows. The entanglement entropy in $2d$ CFT was first calculated by Calabrese et al. \cite{Calabrese:2004eu} and then reproduced by Ryu et al. \cite{Ryu:2006ef} in all dimensions using holographic methods. These are some of the avenues ripe for holographic calculation in asymptotically flat spacetime. \\

\begin{figure}
    \centering
    \includegraphics[width=0.75\linewidth]{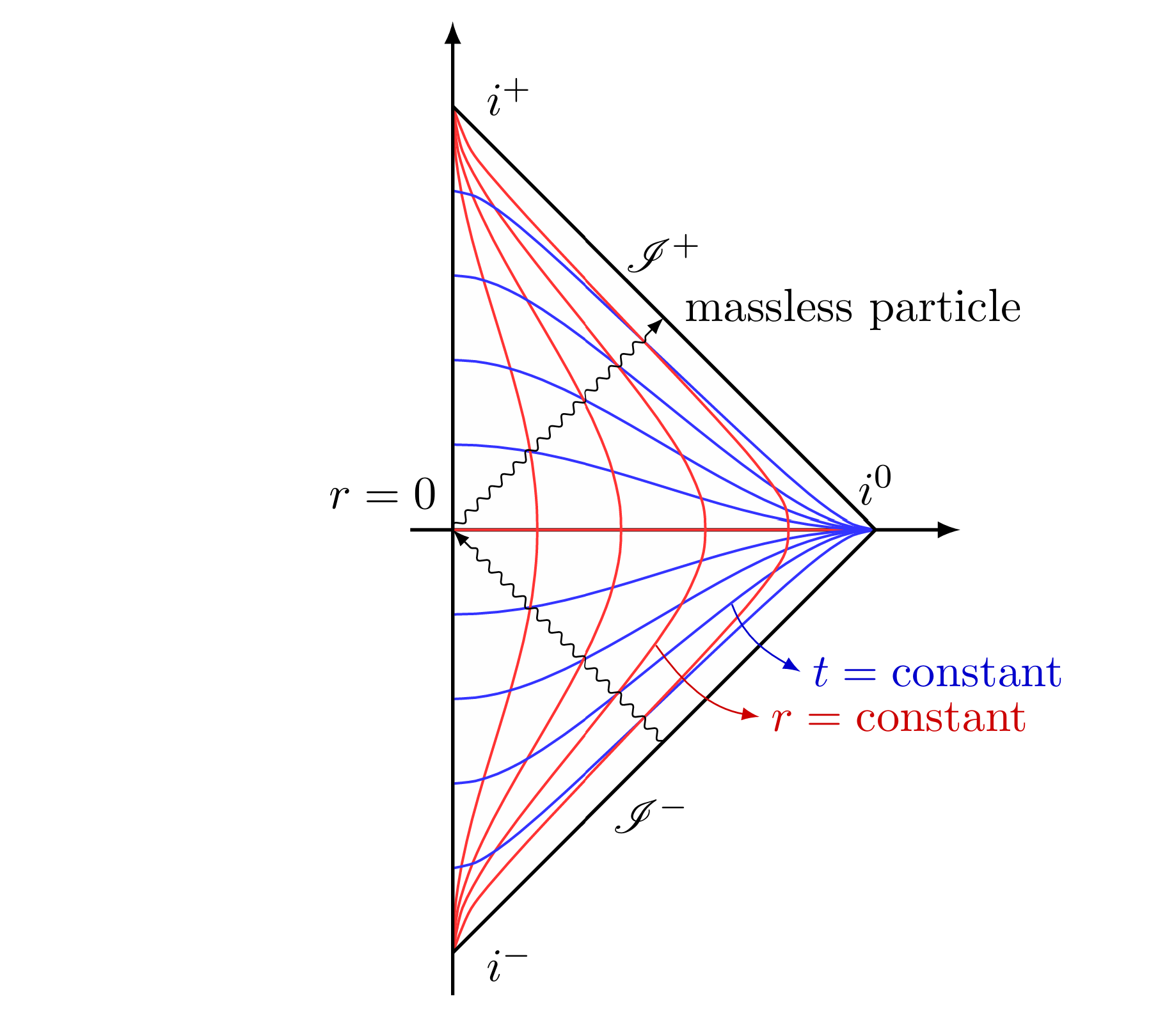}
  \caption{Penrose diagram of 4d Minkowski spacetime with conformal completion. The boundary now consists of five pieces, $i^{-}:$ past timelike inifinity, $i^{+}:$ future timelike infinity, $i^{0}:$ spacelike infinity, $\mathscr{I^{-}}:$ past null infinity, and $\mathscr{I^{+}}:$ future null infinity. Collectively $\mathscr{I}^+ \cup \mathscr{I}^- $ is called the Null boundary, $\mathcal{N}$}
\label{fig:penrose}
\end{figure}

These ideas from AdS can be borrowed and applied to flat spacetime. The conformal structure of flat spacetime (Fig. \ref{fig:penrose}) is more complicated in two important ways compared to AdS spacetime. \\

Firstly, there are two null hypersurfaces and three points, unlike the AdS case, which is simply timelike. Massless particles come from past null infinity, $\mathscr{I}^-$, and go to future null infinity, $\mathscr{I}^+$. While massive particles start from $i^{-}$ and end at $i^{+}$. Finally, there's spatial infinity $i^0$, where one often performs antipodal matching of incoming and outgoing data \cite{Strominger:2017zoo}. In this article, we will focus \textit{only} on the future null infinity, $\mathscr{I}^{+}$ where gravitons and other massless particles finally end up. The analysis for $\mathscr{I^{-}}$ is identical to $\mathscr{I^+}$. The stress tensor at past null infinity $\mathscr{I^{-}}$ can be understood similarly. The stress tensor at $i^{\pm}$ and $i^0$ and its relation to the stress tensor at null infinity are left for future exploration (similar in the spirit of \cite{Ashtekar:2023wfn,Ashtekar:2023zul}).\\

Second, in general relativity, one often needs to add a boundary term to make the variational principle well-defined. For AdS spacetime, the trace of the extrinsic curvature serves the purpose, which is called the Gibbons-Hawking-York term. This term makes the variational principle well defined, but the stress tensor obtained from the variation is often divergent. Balasubramanian et al. \cite{Balasubramanian:1999re} have added suitable counter terms made from the induced metric and its curvature to obtain a finite stress tensor with finite conserved charges. We will follow a similar logic, but in the asymptotically flat case, the boundary is null rather than timelike as is the case for AdS. Hence, for an asymptotically flat spacetime, one requires a different boundary term; and \textit{inaffinity} $\kappa$ fits the bill \cite{Parattu:2015gga,Chandrasekaran:2020wwn,Chandrasekaran:2021hxc,Aghapour:2018icu,Jafari:2019bpw}. Inaffinity is the failure of the null vector to be affinely parameterized. For any arbitrary null vector $n^a$,
\begin{eqnarray}
n^{c}\nabla_c n^a=\kappa\,  n^a
\end{eqnarray}
These two important points make the conformal structure of asymptotically flat spacetime complicated.

Using the well-defined variation principle leads to a candidate expression for the stress tensor\footnote{We defining the mix-indexed stress-energy tensor ${T_j}^i$ to treat it as a map from a vector field $\xi^{j}$ to it's corresponding current $j_{\xi}$ via it's contraction with the volume form $\eta$ as $j_{\xi} = - \eta_{i} {T_j}^i\xi^{j}$. More on this in section 4. }. It is given by the Weingarten tensor and its traces.
\begin{align}
   {T_j}^i &= -\frac{1}{8 \pi G} \left({W_j}^i - W \delta^i_j \right), \quad W = {W_i}^i.
\end{align}

The Weingarten tensor is the projection of the covariant derivative of the normal vector onto the null surface.
\begin{eqnarray}
   {W_j}^i= {\Pi_b}^i(\nabla_a n^b) {\Pi_j}^a
\end{eqnarray}
Projectors ${\Pi_i}^a$ are used to pull back bulk quantities onto the null surface $\mathscr{I}^+$. $\nabla_{a}$ is the standard Levi-Civita connection of the bulk spacetime. We explicitly evaluate the stress tensor \footnote{The stress tensor calculated in this way is finite. Before our article, the stress tensor has been computed using the amplitudes methods \cite{Kapec:2016jld,Ruzziconi:2024kzo,Saha:2023hsl}. The fluid dynamical aspects are analysed in \cite{Bhattacharya:2024fbz,Ciambelli:2023mvj}.} for asymptotically flat spacetime, which has parameters like Bondi mass aspect $M$, angular momentum aspect $N^A$, shear $C_{AB}$, and Bondi news $N_{AB}$. Then we used the stress tensor to compute the conserved charges for global and asymptotic symmetries. Although the charges corresponding to the BMS symmetries have been calculated via different methods like the covariant phase space method by Wald-Zoupas \cite{Wald:1999wa,Barnich:2011mi} and using the Hamiltonian formulation by Ashtaker et al.  \cite{Ashtekar:2024bpi,Ashtekar:2024mme,Ashtekar:2024stm}, we find it useful to find these charges from the Brown-York procedure, which requires \textit{a definition} of the stress tensor. This holographic stress tensor will become the stress tensor of the Carrollian QFT living on the boundary. This is the main motivation for our work. We obtain conserved charges such as mass and angular momentum for Kerr black holes. We have also computed the Brown-York charges for Schwarzschild and the Kerr black hole and correctly reproduced the mass and angular momentum (see section \ref{the Kerr bh} for more details). For angle-dependent symmetries, we found that the charges are compatible with Barnich et al. \cite{Barnich:2011mi}. Further, we calculate the conserved charges for asymptotic BMS \footnote{Bondi-Metzner-Sachs or sometimes also called Bondi-Van der Burg-Metzner-Sachs symmetries.} symmetries. These charges match with the integrable part of the Wald-Zoupas \footnote{For a general Wald-Zoupas formalism see \cite{Wald:1984rg,Wald:1999wa}. The charges with ``relaxed'' boundary condition have been computed in \cite{Ciambelli:2023ott,Geiller:2024amx,Geiller:2024ryw}. }charges calculated by Barnich et.al. \cite{Barnich:2011mi}. Explicitly, we have
\begin{eqnarray}
Q_{BY}=&&\frac{1}{8 \pi G}\int r^2  dz d\bar{z}\Bigg[\left(\frac{2M}{ r^2}\right)({\mathcal{T}+u \alpha})  \nonumber\\
&& +\left(\frac{N_z +\frac{1}{16} \partial_{z}C_{\bar{z}\bar{z}} C_{zz}+\frac{1}{16} C_{\bar{z}\bar{z}} \partial_{z}C_{zz}} { r^2}\right)\mathcal{Y}+ \left(\frac{N_{\bar{z}} +\frac{1}{16}C_{\bar{z}\bar{z}} \partial_{\bar{z}}C_{zz}+\frac{1}{16}\partial_{\bar{z}}C_{\bar{z}\bar{z}} C_{zz}} { r^2}\right)\bar{\mathcal{Y}}\Bigg]+\cdots\nonumber\\
\end{eqnarray}
 Our Brown York charges reproduce the integrable part of their result.\\
 

\textbf{Structure of the article:} In section \ref{AdS}, we review the construction of the holographic stress tensor for AdS spacetime, which will set the stage for the asymptotically flat case. In section \ref{spacetime}, we review the Carrollian structure at null infinity, emphasizing how this structure is induced from the bulk. This guides the choice of a boundary term required for the well-defined variational principle. Then in section \ref{stress tensor}, we use the variational principle to construct a candidate stress tensor suitable for null hypersurfaces in terms of the Weingarten tensor. In section \ref{stress eval}, we compute the stress tensor of asymptotically flat spacetime. Finally, in section \ref{brown}, we evaluate Brown-York charges corresponding to the BMS symmetries and evaluate charges for the Kerr black hole. Some of the discussions about null hypersurfaces are relegated to Appendix \ref{geo null}. \\

\textbf{Main Findings and Contributions of this article:} In this article, we found a stress tensor for an asymptotically flat spacetime in $1/r$ asymptotic expansion. This is analogous to the stress tensor evaluated in \cite{Balasubramanian:1999re} in AdS, which was our primary motivation. This makes our work novel compared to other works in the literature. We foliated spacetime using timelike slices. Using Freidel et al. \cite{Freidel:2022bai,Freidel:2022vjq,Freidel:2024emv} work on induced Carroll structure, we explicitly evaluated the stress tensor and found that it is finite. The respective Brown-York charges for the given BMS symmetries also turn out to be finite. Although the expression for the stress tensor 
\begin{align}
   {T_j}^i &= -\frac{1}{8 \pi G} \left({W_j}^i - W \delta^i_j \right), \quad W = {W_i}^i.
\end{align}
can be found in the literature like \cite{Chandrasekaran:2020wwn,Chandrasekaran:2021hxc,Freidel:2022bai,Freidel:2022vjq,Freidel:2024emv}, an explicit form for the stress tensor for asymptotically flat spacetime with Bondi mass $M$, angular momentum $N_A$, shear $C_{AB}$, and Bondi news $N_{AB}$ has not been done to the best of our knowledge. The evaluation of Brown-York charges for BMS symmetries is also new.  Our concrete construction makes it easier to study and understand the nature of anomalies in asymptotically flat space holography. 
\\

\textbf{Notation:} The beginning letters of the Latin alphabet, $a,b,c,\cdots$ are used to denote the spacetime bulk indices $(4d)$, while the middle letters, $i,j,k,\cdots$ are used for the null hypersurface $(3d)$ indices. Capital letters $A,B$ are reserved as indices for the round metric on $S^2$ or $\mathbb{C}$. Vectors and forms are denoted by unbold \& bold letters, respectively, e.g., vector k, and 1-form \textbf{n}. The spacetime volume form is denoted by $\epsilon$, and for the hypersurface it's $\eta$. Carrollian vector is always defined as $\ell^a$ or $\ell^i$ while the auxiliary null vector is denoted by $k^a$ or $k^i$. There is an additional vector in the bulk, the normal 1-form $n_a$. The $g_{ab}$ represents the bulk $4d$ metric. The metric on $S^2$ is denoted as $q_{AB}$, which is the round metric on the sphere. The degenerate metric on the hypersurface $\mathcal{H}$ is represented as $q_{ab}$. The Carrollian vector $\ell^a$ is in the kernel of $q_{ab}$ as $q_{ab}l^b=0$. At the null infinity, we use middle letters $i,j=\{u,z,\bar{z}\}$ for the boundary coordinates. The induced metric on the hypersurface is denoted by $H_{ab}$. As we will see, the induced metric can be written as
$H_{ab}=q_{ab}- 2\Omega k_a k_b$.



\section{AdS spacetime: a warm-up and a sketch of the formalism}
\label{AdS}
There are numerous ways to foliate a spacetime. In the ADM formalism, (i.e., Hamiltonian formulation of GR), one usually foliates spacetime with a family of spacelike hypersurface  $\Sigma_t$ each at fixed time $t$. One then constructs a Hamiltonian that generates the system's time evolution between the spacelike hypersurfaces $\Sigma_{t} \rightarrow \Sigma_{t+dt}$. In AdS, one usually foliates spacetime with timelike slices at fixed values of the radial coordinate. Pushing one of these slices to infinity ($r \rightarrow \infty$), one reaches the conformal boundary of AdS. Then, one can construct a stress tensor for AdS gravity by varying the gravitational action (along with a boundary term) with respect to the boundary metric. More explicitly
\begin{eqnarray}
\label{stressdef}
T^{\mu\nu}=\frac{2}{\sqrt{-\gamma}} \frac{\delta S_{\mathrm{grav}}[\gamma]}{\delta \gamma_{\mu\nu}}
\end{eqnarray}
where $S_{\mathrm{grav}}[\gamma]$ is the gravitational action as a functional of the boundary metric $\gamma_{\mu\nu}$.\\
The gravitational action for AdS spacetime is
\begin{eqnarray}
S=-\frac{1}{16 \pi G}\int_{\mathcal{M}} d^{d+1} x \sqrt{g} \Bigg(R-\frac{d(d-1)}{l_{AdS}^2}\Bigg)-\frac{1}{8 \pi G} \int_{\partial \mathcal{M}} d^d x \Bigg(\sqrt{-\gamma} K + S_{CT}[\gamma]\Bigg)
\end{eqnarray}
The second term (the trace of extrinsic curvature) makes the variational principle well-defined. In AdS spacetime, counter terms $S_{CT}$ are often needed to make the stress tensor and conserved charges finite. Upon varying the action \cite{Balasubramanian:1999re} we get,
\begin{eqnarray}
T^{\mu\nu}=\frac{1}{8 \pi G} \Bigg(K^{\mu\nu}- K \gamma^{\mu\nu}+\frac{2}{\sqrt{-\gamma}} \frac{\delta S_{CT}}{\delta \gamma_{\mu\nu}}\Bigg)
\end{eqnarray}
As an example, for AdS$_3$ gravity, the counter terms and stress tensor can be written as
\begin{eqnarray}
S_{CT}=\frac{1}{8 \pi G}\int_{\partial M} d^d x\,  \left(\frac{-1}{l_{AdS}}\right) \sqrt{\gamma} \implies  T^{\mu\nu}=\frac{1}{8 \pi G} \Bigg(K^{\mu\nu}- K \gamma^{\mu\nu}-\frac{1}{l_{AdS}} \gamma_{\mu\nu}\Bigg)
\end{eqnarray}
Similar counter terms and stress tensors for higher-dimensional AdS spacetime can be explicitly written. The stress tensor depends on the geometric structure (intrinsic as well as extrinsic) of the boundary metric. Then AdS/CFT correspondence asserts the above stress tensor as an expectation value of the CFT stress tensor.
\begin{eqnarray}
\langle T^{\mu\nu} \rangle =\frac{2}{\sqrt{-\gamma}} \frac{\delta S_{\mathrm{grav}}}{\delta \gamma_{\mu\nu}}
\end{eqnarray}

This relation connects the stress tensor computed using e.q. \ref{stressdef} with the CFT stress tensor, making the former \textit{holographic}. In the above analysis, it was crucial that the authors of \cite{Balasubramanian:1999re} foliated the spacetime by constant $r$ slices and then took $r \rightarrow \infty$ limit. In this limit, the slice can be identified with the boundary.

\section{Asymptotically flat spacetime:}
\label{spacetime}
Asymptotically flat spacetime is a solution to Einstein's equations with zero cosmological constant,
\begin{equation}
\label{einstein equations}
G_{ab} \equiv R_{ab}-\frac{1}{2} g_{ab} R=0.
\end{equation}
The Penrose diagram of flat spacetime is shown in fig \ref{fig:penrose}. The boundary $\mathscr{I}^+$ is null. We foliate the spacetime by constant $r-$slices \footnote{There is another choice: the constant $v= t+r$ slices. These slices are null at any constant $v$. } (timelike) as shown by red lines in fig \ref{fig:penrose}. When one takes $r \rightarrow \infty$ limit these slices can be identified with $\mathcal{N} \equiv \mathscr{I}^+ \cup \mathscr{I}^- $. Our primary interest will be in the future null infinity $\mathscr{I}^+$.
\subsection{Carrollian structure at null boundary:}
\label{structure}
A general solution of Einstein's equations can be written as an asymptotic expansion in Bondi gauge as (for the most general solution see section \ref{sol Einstein})
\begin{eqnarray}
ds^2&&= \Bigg(-1+\frac{2 M}{r}+O(r^{-2})\Bigg) du^2- 2 (1+O(r^{-2})) du dr+\Big(r^2 q_{AB} +r C_{AB} +O(r^{0})\Big)dx^A dx^B\nonumber\\
&& +2\Bigg(\frac{1}{2} D_B {C_A}^B +\frac{2}{3 r} (N_A+\frac{1}{4} {C_A}^B D_C {C_B}^C)+O(r^{-2})\Bigg) du dx^A +\cdots
\end{eqnarray}
At null infinity, since $dr=0$, and we pull the conformal factor $r^2$, then we have \footnote{The $\hat{=}$ in the expression below means the equality only at the null infinity.}
\begin{eqnarray}
d\tilde{s}^2&&=\frac{1}{r^2}\Bigg(-1+\frac{2 M}{r}\Bigg) du^2+\Big( q_{AB} +\frac{1}{r} C_{AB} \Big)dx^A dx^B
\\ &&+\frac{2}{r^2}\Bigg(\frac{1}{2} D_B {C_A}^B +\frac{2}{3 r} (N_A+\frac{1}{4} {C_A}^B {D_C C_B}^C)\Bigg) du dx^A  \nonumber\\
&& \hat{=} \quad q_{AB} dx^A dx^B
\end{eqnarray}
Here $q_{AB}dx^A dx^B= \frac{4}{(1+z \bar{z})^2} dz d\bar{z}$ is the metric on the $S^2$, with $ (z,\bar{z}) \in \mathbb{C}$. The weak \footnote{in contrast a strong Carroll structure consists of a connection $\nabla$ and possibly a compatible metric along with $q_{ij}$ and $n^i$ } Carroll structure consists of a pair $(q_{ij},n^i)$ on the manifold which describes the null boundary, where $q_{ij}$ is a degenerate metric and $n^i$ is a preferred direction (vector field) chosen to be in the kernel of the metric as $q(n, \phantom{.}) = 0$. For Minkowski spacetime (set all parameters $m_{B}, C_{AB}, N_A, N_{AB}$ in the above metric to zero), the preferred direction is $n^a \partial_a= \partial_u$. In the subsequent sections, we discuss how the Carroll structure gets induced from the bulk spacetime.


\subsection{Induced Carroll structure on constant $r$ hypersurface}
\label{indCarll}
First, we review the formalism of induced Carroll structure on fixed $r$ hypersurface in $4d$ bulk spacetime \cite{Freidel:2022bai,Freidel:2022vjq}. Next, we see how the Carroll structure on this hypersurface gets identified with the Carroll structure at null infinity $\mathcal{N}$ when the hypersurface is pushed to the boundary. \footnote{The formalism works for any null hypersurface $\mathcal{N}$ but we are interested in null infinity.}\\  

We start with a 4d spacetime $\mathcal{M}$ with a Lorentz signature metric $g_{ab}$ and the Levi-Civita connection $\nabla_{a}$. We foliate this spacetime by a family of 3-dimensional timelike slices at fixed $r$, called \textit{stretched Horizon(s)} $\mathcal{H}$ \cite{Riello:2024uvs,Freidel:2024emv}. $\mathcal{H}$ is parametrized by $r(x)=r_0(u,z,\bar{z}) >0$. Next, we equip $\mathcal{H}$ with an extrinsic structure called the \textit{rigged structure} (explained below). The Carroll structure is then induced from the rigged structure and more importantly it becomes an intrinsic structure on $\mathcal{H}$. After pushing the slice to the boundary, this Carroll structure gets identified with the Carroll structure at  $\mathcal{N}$ (see Donnay et al.\cite{Donnay:2019jiz} for a similar construction at the black hole horizon). Let's discuss some definitions.\\

\begin{figure}
\centering
\includegraphics[width=1\linewidth]{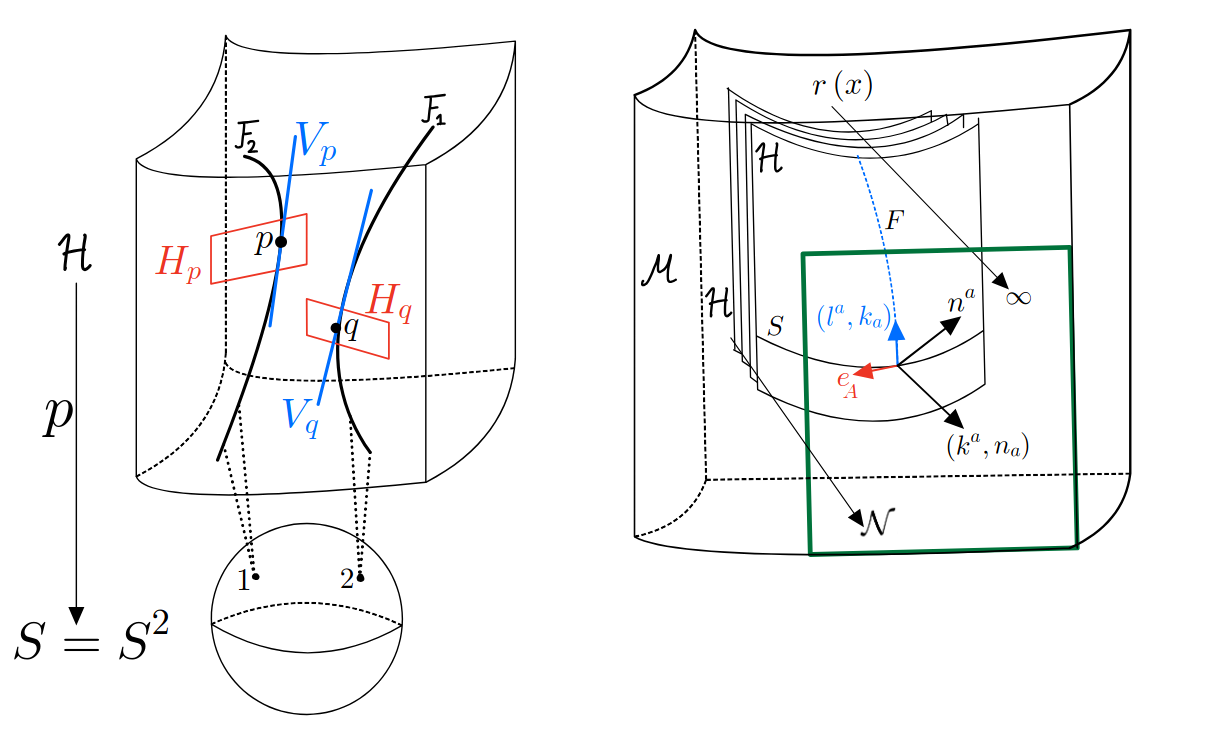}
\caption{\textbf{Left Panel:} Depicts a single fiber bundle $p:\mathcal{H} \rightarrow \mathcal{S}$. $\mathcal{F_{1}}$ is an example of a 1-dimensional fiber that is attached to a point on the base space, $\mathcal{S} = S^2$. $V_{p} \equiv \text{Ver}_{p}(T\mathcal{H})$ and $H_{p} \equiv \text{Hor}_{p}(T\mathcal{H})$ are the vertical 1-dimensional (blue) and horizontal 2-dimensional (red) disjoint vector subspaces of $TH$. \textbf{Right Panel:} Stack of stretched horizons, $\{H\}$ embedded in $\mathcal{M}$. Here we suppress one dimension to better emphasize how various vectors are aligned. The Carrolian vector and the Ehresmann connection, ($\ell^a, k_a$), point along the fiber (blue). Basis vectors $e_{A}$ (red) on $\mathcal{S}$ lie in $TS$. The rigged structure ($k^a, n_a$) points along $r(x)$. Also, note that the normal vector $n^a \rightarrow \ell^a$ as $H \rightarrow \mathcal{N}$. Note that the described orientation of various vectors and 1-forms remains valid as long as the background geometry is static. If we have a dynamic background (e.g., propagating graviton), then these orientations would change. We will see some examples of these decompositions in section \ref{stress eval}. Note that the above cartoon is color coded to cleanly represent various parts of the geometry and the relevant geometric objects within them.}
\label{fig:rigged}
\end{figure}

\textbf{Carroll structure:} The \textit{Carrollian Viewpoint} consists of splitting the tangent space $T_{p}\mathcal{H}$ (at point p) of the stretched horizon $\mathcal{H}$ into vertical $V_{p} \equiv \text{Ver}(T_{p}\mathcal{H}) $ \& horizontal $H_{p} \equiv \text{Hor}(T_{p}\mathcal{H}) $ subspaces \footnote{This is useful when this slice is pushed to $\mathcal{N}$ where this decomposition helps to isolate the null direction in the $\text{Ver}(T\mathcal{H})$ subspace, from the remaining directions in $\text{Hor}(T\mathcal{H})$.} (See fig \ref{fig:rigged} for more details) so that the entire tangent bundle $T\mathcal{H}$ decomposes as
\begin{equation}
TH = \text{Ver}(T\mathcal{H}) \bigoplus \text{Hor}(T\mathcal{H})  
\end{equation}

We begin by treating the stretched horizon $\mathcal{H}$ as a line bundle $p:\mathcal{H} \rightarrow \mathcal{S}$, where the base space $\mathcal{S} = S^2$ and the 1-dimensional fiber $\mathcal{F}$ is attached to each point on $\mathcal{S}$. $\mathcal{F}$ is spanned by a \textit{Carrollian vector} $\ell \in ker(\textbf{d}p:T\mathcal{H} \rightarrow T\mathcal{S}$) (which is tangent to and points along the fibre $\mathcal{F}$ see Fig. \ref{fig:rigged} right panel). Thus $\ell$ defines the $\text{Ver}(T\mathcal{H})$. Next, we introduce the \textbf{Ehresmann connection} 1-form \textbf{k} which helps in defining $\text{Hor}(T\mathcal{H})$ by constructing a basis ${e_A}$ \footnote{Remember that A is an index on $S^2$ hence runs over {1,2} or {$z,\bar{z}$}} with respect to \textbf{k} such that $\iota_{e_{A}}\textbf{k} = 0$. Basis $e_{A}$ spans $\text{Hor}(T\mathcal{H})$. By construction, the Ehresmann connection 1-form \textbf{k} is dual to the Carrollian vector $\ell$ i.e. $\iota_\ell \textbf{k} = -1$. Next, we can chart $\mathcal{S}$ with local coordinates $\{\sigma^A\}$ and a round metric $q_{AB}$. The \textit{null Carrollian metric} on $\mathcal{H}$ is defined as a pull-back of the round metric $q_{AB}$ onto $\mathcal{H}$ i.e. $q = p^*(q_{AB} d\sigma^A d\sigma^B)$ making $\ell$ its kernel i.e. $q(\ell, .) = 0$. The collection $(\mathcal{H}, q, \ell, \textbf{k})$ is called the \textbf{Carroll structure.}\\

\textbf{Null rigged structures:} To understand the geometry of $\mathcal{H}$, we use the rigging technique developed by Mars and Senovilla \cite{Mars_1993}. This involves embedding a any level set of hypersurface (which later become the stretched horizons $\{\mathcal{H}\}$) in spacetime $\mathcal{M}$ using the foliation $r(x) =r_0$  $>0$. Due to embedding, there is a natural notion of a normal 1-form, $\mathbf{n}= n_a dx^a$, where $n_a = \partial_{a}r(x)$. A \textit{rigging vector} field, $k=k^a \partial_a$ is then introduced which is transverse \footnote{transverse in this context means nowhere tangent to the stretched horizon, $\mathcal{H}$ and not necessarily normal $\mathcal{H}$, rather it's normalization is fixed with respect to \textbf{n}. This construction may seem trivial for a timelike surface, but once this rigged structure becomes null, $k$ will be identified with the auxiliary vector field needed for null hypersurfaces, as explained in the appendix.} $\mathcal{H}$ and is dual to the normal 1-form, $\iota_{k}\textbf{n} = -1$. The pair of dual rank-1 tensors $(\textbf{n}, k)$ is known as the \textbf{rigged structure} (fig. \ref{fig:rigged} right panel).\\

Now we introduce a projection operator that will project all tensors defined in $\mathcal{M}$ into $\mathcal{H}$. The \textit{rigged projection tensor}, $\Pi: T\mathcal{M} \rightarrow T\mathcal{H}$ is defined as
\begin{eqnarray}
{\Pi_a}^b= {\delta_a}^b + n_a k^b, \quad k^a {\Pi_{a}}^b = {\Pi_a}^b n_b = 0
\end{eqnarray}
As an example, if $X \in T\mathcal{M}$ and $\boldsymbol{\omega} \in T^{*} \mathcal{M}$, then $\overline{X}^b := X^a {\Pi_a}^b \in T\mathcal{H}$ and $\overline{\omega}_a := {\Pi_{a}}^{b}\omega_b  \in  T^{*}\mathcal{H}$. These projected tensor satisfies $\overline{X}^a n_a = k^a \overline{\omega}_a = 0$.\\

To treat timelike and null surfaces simultaneously, one often chooses the rigging vector to be null by requiring $\overline{k}_a:=k_a$ \cite{Freidel:2022vjq}, where $k_a=g_{ab}k^b$. Thus, we have $\overline{k}_a k^a = k_{a}k^{a} = 0$ making $k$ a \textit{null rigging vector},
\begin{eqnarray}
k_a k^a=0, \quad \text{while}\quad n_a n^a= 2 \Omega, \quad  \text{and} \quad \Omega \rightarrow 0 \text{ at the null infinity}
\end{eqnarray}
Thus, the pair $(\textbf{n},k)$ with $k$ being a null vector, defines the \textbf{null rigged structure} on $\mathcal{H}$. The final piece of ingredient is a tangential vector field  $\ell= \ell^a \partial_a \in TH$ and whose components are given by,
\begin{eqnarray}
\label{norm}
\ell^a := n^b {\Pi_b}^a \implies \quad \ell^a= n^a + 2\Omega k^a, \quad \ell_a \ell^a= -2\Omega
\end{eqnarray}

One can check that $\iota_{\ell} \textbf{n}=0 \text{ and} \quad \iota_{\ell} \textbf{k}=-1$. With these properties, it is evident that $\ell$ could be thought as of a Carrollian vector \footnote{Note that even though we introduced $\ell$ while constructing the null rigged structure and later identified it with the Carrolian vector, it's not a null vector in the embedding space, $\mathcal{M}$ (as opposed to $k$ which is a null vector in $\mathcal{M}$), rather $\ell$ is a null vector strictly on $\mathcal{N}$.} and $k_a = g_{ab}k^b$ becomes the Ehresmann connection. The normalization $\Omega$ from \eqref{norm} captures the idea of \textit{stretching} of H, in the spirit that if $ \Omega \neq 0$ then the metric defines $\mathcal{H}$ as a timelike surface while when $\Omega=0$ the metric defines $\mathcal{H}$ as a null surface. As $H \rightarrow \mathcal{N}$, $\Omega \rightarrow 0$ hence $n^a \rightarrow \ell^a$. With the help of $\ell$ and $\textbf{k}$ when needed, we can decompose the rigged projection tensor into horizontal and vertical bases to provide the Carrollian viewpoint
\begin{eqnarray}
{\Pi_a}^b= {q_a}^b + k_a \ell^b, \quad {q_a}^b k_b=0=\ell^a {q_a}^b.
\end{eqnarray}
Here ${q_a}^b={e_a}^A {e_A}^b$ is the projector onto the horizontal subspace i.e. ${q_a}^b : T\mathcal{H} \rightarrow \text{Hor}(T\mathcal{H})$. The degenerate metric on $\mathcal{H}$ can be constructed by $q_{ab}={q_a}^c {q_b}^d g_{cd}$ and further restricting it on $\mathcal{H}$ i.e. $q_{ab} \rightarrow q_{ij}$ . This is an important result, hence we will summarize what we have done so far.\\

\underline{\textbf{Result:}} Given a null rigged structure $(\textbf{n},k)$ on an embedded stretched horizon $\mathcal{H}$ and the embedding spacetime metric $g$ the Carroll structure $(\mathcal{H},q, \ell, \textbf{k})$ gets naturally induced on $\mathcal{H}$ where
\begin{eqnarray}
\ell^a = n_b g^{bc}{\Pi_c}^a, \quad k_a = g_{ab}k^b, \quad \text{and} \quad q_{ab} = {q_a}^c {q_b}^d g_{cd}  
\end{eqnarray} 
The vectors $(\ell^a,k^a,{e_A}^a)$ span the tangent space $T\mathcal{M}$ while 1-forms $(k_a, n_a, {e_a}^A)$ span the cotangent space $T^{*}\mathcal{M}$. The spacetime metric $g_{ab}$ is decomposed into the horizontal, vertical, and radial directions as 
\begin{eqnarray}
\label{withOmega}
\nonumber g_{ab} &=& q_{ab}-k_a \ell_b - n_a k_b\\
&=& q_{ab} - 2 n_{(a}k_{b)} - 2\Omega k_a k_b
\end{eqnarray}
 Next, the \textbf{rigged metric} and its dual can be written with the help of the projection tensor
\begin{eqnarray}
H_{ab}= {\Pi_a}^c {\Pi_b}^d g_{cd}, \quad H^{ab}= g^{cd} {\Pi_c}^a {\Pi_d}^b,\quad H_{ab}k^b=0\\
H_{ab}= q_{ab}- 2\Omega k_a k_b, \quad H^{ab}=q^{ab}
\end{eqnarray}
A \textit{rigged connection} on $\mathcal{H}$ is simply the projection of the Levi-Civita connection $\nabla$ on  $\mathcal{H}$ from $\mathcal{M}$ 
\begin{eqnarray}
D_{a}{A_b}^{c}={\Pi_a}^d {\Pi_b}^e (\nabla_d {A_e}^f) {\Pi_f}^c
\end{eqnarray}
here ${A_b}^c$ is any tensor field defined on $\mathcal{M}$. This rigged connection is torsion-free and preserves the rigged projection tensor (see section 2.5 of \cite{Freidel:2022vjq} for more details.)
However, the rigged metric is not covariantly constant with respect to the rigged connection. Instead, it is proportional to the extrinsic curvature of $\mathcal{H}$
\begin{eqnarray}
D_a H^{bc}= -({K_a}^b \ell^c+ {K_a}^c \ell^b)
\end{eqnarray}
where ${K_a}^b={\Pi_a}^c  \nabla_ck^d {\Pi_d}^b$ is the extrinsic curvature, sometimes called the \textit{rigged extrinsic curvature}. In similar spirit, the volume form $\eta$ on $\mathcal{H}$ can be obtained from the volume form $\epsilon$ of spacetime $\mathcal{M}$ by contraction with the null rigging vector i.e. $\eta:= \iota_k \epsilon$.\\

\subsection{Intrinsic and extrinsic geometry at null infinity}\label{3}
We have established that one can induce a Carrollian structure from the null rigid structure defined on $\mathcal{H}$. Moreover, this structure persists as we take $r \rightarrow \infty$ i.e $\mathcal{H} \rightarrow \mathcal{N}$.\\

Let's first understand the intrinsic geometry of the null infinity, which has coordinates $x^i$ with indices $i=\{u,z,\bar{z}\}$, and $n^{a}=\ell^{a}$. At null infinity, the geometry consists of a degenerate metric $q_{ij}$ and a preferred vector $n^i$ which is in the kernel of metric i.e., $n^i q_{ij}=0$. These objects define a weak Carroll structure on the null infinity $\mathcal{N}$. The volume form on horizontal space (which has metric $q_{AB}$) is labeled by $\mu$ and the total volume form on $\mathcal{N}$ is denoted as $\eta$ and satisfies the following relations.
\begin{eqnarray}
d \mu = \Theta \eta ,  \quad K_{ij}= \frac{1}{2} \mathcal{L}_{n} q_{ij}, \quad 
\sigma_{ij}= K_{ij}-\frac{1}{2} \Theta q_{ij}
\end{eqnarray}

Here $\Theta \equiv q^{ij}K_{ij}$ is called the expansion scalar, $K_{ij}$ is the extrinsic curvature (although completely determined by intrinsic data $(q_{ij},n^i)$), and $\sigma_{ij}$ is the shear. The Weingarten tensor \footnote{Also known as the Shape tensor} ${W_j}^i$ of the null boundary $\mathcal{N}$ describes the extrinsic data. It can be written as
\begin{eqnarray}
{W_j}^i = {\Pi_j}^a \nabla_a n^b {\Pi_b}^i
\end{eqnarray}

Here we have used a mixed index projection tensor which is useful for projecting the spacetime tensors (with index $a=1,2,3,4$) onto $\mathcal{N}$, which has index as $i=1,2,3$
\begin{eqnarray}
{\Pi_i}^a= {\delta_i}^a+n_i k^a, \quad {\Pi_a}^i= {\delta_a}^i+n_a k^i, \quad \text{and}\,\, {\Pi_b}^a= {\delta_b}^a+ n_b k^a= {\Pi_b}^i {\Pi_i}^a.
\end{eqnarray}

The Carrollian vector and the Ehresmann connection on $\mathcal{N}$ are $n^i= {\Pi_a}^i n^a$ \footnote{as $H \rightarrow \mathcal{N} \implies n^{a} \rightarrow \ell^{a}$. But we keep using the $n^i$ as Carrollian vector on null infinity.} and $k_i={\Pi_i}^a k_a$ respectively.\\

The Weingarten tensor satisfies the following identities
\begin{eqnarray}
{W_j}^i q_{ik} =K_{jk}, \quad {W_j}^i n^j= \kappa n^i
\end{eqnarray}
The preferred vector $n^i$ is the eigenvector of the Weingarten tensor with eigenvalues $\kappa$ called \textbf{inaffinity}. This inaffinity will become the relevant boundary term for a well-defined variational principle for gravitational action as we see in the next section. The vector $n^i$ is null and hence generates the null geodesic on $\mathcal{N}$ which satisfies the geodesic equation as
\begin{eqnarray}
n^j \nabla_j n^i= \kappa \, n^i
\end{eqnarray}
As before, the projection tensor can be used to define a rigged connection on $\mathcal{N}$ from the spacetime Levi-Civita connection \cite{Mars_1993} as
\begin{eqnarray}
D_i {A}^j= {\Pi_i}^a {\Pi_b}^j D_a A^b, \quad \text{where}\, \, D_a A^b= {\Pi_a}^c {\Pi_d}^b \nabla_c A^d
\end{eqnarray}
This connection is torsionless but not \textit{compatible} as
\begin{eqnarray}
\label{connection}
D_i q_{jk}=k_j K_{ik}+k_k K_{ij},\quad  D_i \eta= - \rho_i \eta\\
D_i n^j= {W_i}^j={K_i}^j+ \rho_i n^j, \quad D_i \mu= {K_i}^j \eta_j
\end{eqnarray}
The connection can also act on covariant vectors as
\begin{eqnarray}
D_i A_j= {\Pi_i}^a {\Pi_j}^b \nabla_a( {\Pi_b}^k A_k)
\end{eqnarray}
The auxiliary vector $k_i$ relative to $n^i$ such that $n^i k_i=-1$. With this one form, one can also define a projector on the horizontal forms as
\begin{eqnarray}
{q_j}^i={\delta^i}_j+k_j n^i 
\end{eqnarray}
This also helps us in defining the inverse metric as $q_{ij}q^{jk}={q_{i}}^k.$ The shape operator can be decomposed as
\begin{eqnarray}
{W_j}^i={K_j}^i+\rho_j n^i , \quad \rho_j=\bar{\omega}_j- \kappa k_j
\end{eqnarray}
Here ${K_j}^i= q^{ik} K_{kj}$ and $\rho_j$ is called rotation one form which is defined by
\begin{eqnarray}
\rho_j =- {\Pi_j}^a k_b \nabla_a n^b, \quad \bar{\omega}_i= {q_i}^j \rho_j
\end{eqnarray}
here $\bar{\omega}$ is  H\'{a}j\'{i}\v{c}ek one form. 

\section{Stress tensor}
\label{stress tensor}
In this section, we obtain the expression for the holographic stress tensor from the variation of the gravitational action.

\subsection{Gravitational action and variational principle}
The gravitational action with the boundary term is given
\begin{eqnarray}
\label{grav_action}
S=\frac{1}{16 \pi G} \int_{\mathcal{M}} d^4 x \sqrt{g} R+\frac{1}{8 \pi G} \int_{\mathcal{N}} \eta \, \kappa
\end{eqnarray}
In the second integral, $\kappa$ is the inaffinity while $\eta$ is the invariant volume form on the boundary. The variation of the action \eqref{grav_action} with respect to the Carroll structure i.e. boundary metric $q_{ij}$ and the normal vector $n^i$ is
\begin{eqnarray}
\label{variationS}
\delta S&&=  \int d^4 x \sqrt{g} E^{ab} \delta g_{ab}+\int  \Big(\pi^{ij}\delta q_{ij}+\pi_i \delta n^i- d\beta\Big)\nonumber\\
\pi^{ij}&&= -\frac{1}{16 \pi G} (K^{ij}- (\Theta+\kappa) q^{ij})\eta\nonumber\\
\pi_i&&= \frac{1}{8 \pi G} (\Theta n_i+\bar{\omega}_i)\eta\nonumber\\
\beta&&= \frac{1}{16 \pi G} (g^{ab}n^c-n^a g^{bc}) \delta g_{ab} \eta_c\nonumber\\
\end{eqnarray}
The variation of the action is zero whenever the equation of motion are satisfied and the Carroll structure  $(q_{ij}, n^i)$ is fixed \cite{Chandrasekaran:2020wwn,Parattu:2015gga}. Hence on-shell classical action should be thought of as a functional of Carroll structure  $S[q_{ij},n^i]$.
Then the variation of the action with vector parameter $\xi$ which changes the boundary metric as $\delta_{\xi} q_{ij}= \mathcal{L}_{\xi} q_{ij}$ and normal vector as $\delta_{\xi} n^i= \mathcal{L}_{\xi} n^i$ is
\begin{eqnarray}
&& \delta_{\xi} S= - \int_{\mathcal{N} }\mathcal{E}_{\xi}, \quad \text{where}\, \\
&& \mathcal{E}_{\xi}= - \pi^{ij}(\xi^k D_k q_{ij}+2 D_i \xi^k q_{kj})- \pi_i (\xi^k D_k n^i-n^k D_k \xi^i)
\end{eqnarray}
The expressions above can be rearranged as
\begin{eqnarray}
\mathcal{E}_{\xi}= d j_{\xi}- f_{\xi}
\end{eqnarray}
The choice of $j_{\xi}, f_{\xi}$ can be made unambiguously. This decomposition defines the stress tensor by $j_{\xi}\equiv-{T_j}^i \xi^j \eta_i$ and $f_{\xi}= - (\mathrm{div} \, T)_j \xi^j \eta$. In terms of undensitized momenta (excluding the volume form), which is defined as $\pi^{ij}= p^{ij} \eta, \, \pi_i= \eta \, p_i$, the stress tensor and generalized divergence are
\begin{eqnarray}
\label{tij}
&&{T_j}^i= 2  q_{jk} p^{k i}-  p_j n^i\\
&& (\mathrm{div}\, T)_j= D_i {T_j}^i-  {T_j}^i \rho_i- D_j q_{ik}  p^{ik}-  D_j n^ip_i
\end{eqnarray}
The above expression for the stress tensor can now be written in terms of the Weingarten tensor as
\begin{eqnarray}
{T_j}^i= -\frac{1}{8 \pi G}({W_j}^i- W {\delta_j}^i), \quad \quad {W_j}^i= {\Pi_b}^i(\nabla_a n^b) {\Pi_j}^a
\end{eqnarray}
This is the analog of the stress tensor obtained in AdS space using extrinsic curvature (see section \ref{AdS} for a summary). In the flat space (with a null boundary), the \textit{shape} operator (Weingarten tensor) plays the role of extrinsic curvature. The $W^i_j$ is independent of the choice of auxiliary vector $k_i$, and then the mixed index tensor $T^i_j$ is independent of $k_i$. Other index tensors like $T^{ij}$ or $T_{ij}$ might depend on the choice of $k_i$. The stress tensor can be shown to be covariantly conserved on the boundary \cite{Chandrasekaran:2021hxc}\footnote{The divergence of stress tensor is $(\mathrm{div} T)_j=D_i {T_j}^i- {T_j}^i \rho_i- D_j q_{ik} p^{ik}-D_j n^i p_i$. The extra terms other than $D_i {T^i}_j$ cancel. The connection $D_i$ is a rigged connection that is torsionless but not metric compatible. }
\begin{eqnarray}
( \mathrm{div }T)_j= D_i T_j^i=0
\end{eqnarray}
The first equality is obtained using the eq. \eqref{connection}. The second equality above can be obtained using the Codazzi equation for the null surfaces along with the vacuum solution of Einstein's equations\footnote{The simplification in the second equality of Eq. (4.9) uses the vacuum Einstein equations $R_{ab}=0$ to eliminate curvature contributions appearing in Codazzi-type relations. Thus, this step assumes the on-shell condition, not just geometric identities.} \cite{Gourgoulhon:2005ng,Chandrasekaran:2021hxc}. It is also established in \cite{Chandrasekaran:2021hxc} that when the symmetry transformation acts \textit{covariantly} on the boundary, the Brown-York charges (calculated using the stress tensor), match with the ones calculated from Wald-Zoupas procedure. 
\subsection{Solutions of Einstein equation}
\label{sol Einstein}
Einstein's equation in flat spacetime is $G_{ab}=R_{ab}-\frac{1}{2}g_{ab} R=0$. A general solution of Einstein's equation in Bondi gauge with coordinates as $(u,r,x^A=(z,\bar{z}))$ has the following asymptotic expansion in the large $r$ limit \cite{Donnay:2022aba,Donnay:2022wvx,Strominger:2017zoo,Barnich:2011mi,Barnich:2010eb}.
\begin{eqnarray}
    ds^2=e^{2 \beta} \frac{V}{r} du^2-2 e^{2 \beta} du dr+g_{AB}(dx^A-U^A du)(dx^B-U^B du)
\end{eqnarray}
Here, the various parts of the above metric are \footnote{In the Bondi-Sachs metric, \(\beta\) is a scalar function encoding radial/angular warp factors appearing in the full 4D metric expansion, whereas \(\phi\) is the conformal factor that parametrizes the geometry of the 2-dimensional angular metric \(\bar{\gamma}_{AB} = e^{2\phi} \tilde{\gamma}_{AB}\) on the sphere at null infinity. }

\begin{eqnarray}
    g_{AB}=r^2 \bar{\gamma}_{AB}+ r \, C_{AB} +\frac{1}{4}q_{AB} C^C_DC^D_C, \quad \bar{\gamma}_{AB}dx^A dx^B= e^{2 \phi}(d \theta^2+ \mathrm{sin}^2 \theta \, d\phi^2)
\end{eqnarray}
$C_{AB}$ is a symmetric traceless tensor, and other tensors are raised and lowered by the $\bar{\gamma}_{AB}$ and its inverse. The news tensor is defined as $N_{AB}= \partial_u C_{AB}$ and 
\begin{eqnarray}
    &&\beta= -\frac{1}{32 r^2} C_{B}^A C^B_A+O(1/r^3), \quad \frac{V}{r}= - \frac{1}{2} \bar{R}+\frac{2 M}{r}+O(1/r^2)\\
    && U^A= -\frac{1}{2 r^2} \bar{D}_B C^{BA}-\frac{2}{3 r^3}\left[-\frac{1}{2}C_B^A \bar{D}_C C^{BA}+N^A\right]+O(1/r^{4})
\end{eqnarray}

Then we have $g_{uA}$ as\footnote{Often in $U^A$ and $g_{uA}$ there is a logarithmic term like $\frac{log\, r}{r}$ and its coefficients are $D_{AB}$. But here we are not including those terms. See \cite{Barnich:2011mi,Barnich:2010eb} for more details. }
\[
    g_{uA}= -\frac{1}{2 } \bar{D}_B C^{B}_A+\frac{2}{3 r}[-\frac{1}{2}C_{AB} \bar{D}_C C^{CB}+N_A]+O(1/r^{2}) \footnotemark \\   
\]

Here  $\bar{D}$ is the covariant derivative with respect to $\bar{\gamma}_{AB}$ and $N_A$ is called the angular momentum aspect. The Ricci scalar is 
\begin{eqnarray}
    \bar{R}= 2 e^{-2 \phi}- 2 \bar{\Delta} \phi.
\end{eqnarray}
Here $\bar{\Delta}$ is the Laplacian of the metric $\bar{\gamma}_{AB}$. The Einstein equation $E_{uu}=0$ and $E_{uA}$ are constraint equations and give the evolution of mass and angular momentum aspects.
\begin{eqnarray}
    &&\partial_u M= -\frac{1}{8}  N^B_A N_B^A+\frac{1}{8} \bar{\Delta} \bar{R}+\frac{1}{4} \bar{D}_A \bar{D}_C N^{CA}\\
    &&\partial_u N_A= \partial_A M+\frac{1}{4} C_A^B \partial_B \bar{R}+\frac{1}{16}\partial_A[C^C_BN_C^B]-\frac{1}{4} \bar{D}_A  N_C^B C^C_B\nonumber\\
    &&-\frac{1}{4} \bar{D}_B[ C_C^B N^C_A-N_C^B C^C_A]-\frac{1}{4} \bar{D}_B[\bar{D}^B \bar{D}_C C^C_A-\bar{D}_A \bar{D}_C C^{BC}]
\end{eqnarray}
Hence, the solution space  is parametrized by
\begin{eqnarray}
    \mathcal{X}^{\Gamma}\equiv \{C_{AB},N_{AB},M,N_A\}
\end{eqnarray}
The metric on the boundary is often taken to be the round 2 sphere metric.
\begin{eqnarray}
    \bar{\gamma}_{AB}dx^A dx^B= \frac{4}{(1+z \bar{z})^2} dz d\bar{z}, \quad \text{then} \quad \bar{R}=2.
\end{eqnarray}
In this article, we will take the flat metric on the boundary as
\begin{eqnarray}
    \bar{\gamma}_{AB}dx^A dx^B= 2 dz d\bar{z}, \quad \text{then} \quad \bar{R}=0.
\end{eqnarray}
This parametrization is often useful when doing explicit calculations. However, similar conclusions can be reached for other metrics as well. Let's first take the example of Minkowski spacetime in this parametrization. We start with the following redefinition of coordinates as
\begin{eqnarray}
\label{weyl mink}
r=\frac{\sqrt{2}}{1+z_o\bar{z}_o}r_o+\frac{u_o}{\sqrt{2}}, \quad u= \frac{1+z_o \bar{z}_o}{\sqrt{2}}u_o-\frac{z_o \bar{z}_o u_o^2}{2 r}, \quad z=z_o- \frac{z_o u_o}{\sqrt{2} r}
\end{eqnarray}

Here $(u_o,r_o,z_o,\bar{z}_o)$ all are usual Bondi coordinates. One can also relate this new coordinate to the Cartesian ones as
\begin{eqnarray}
X^{\mu}= u \partial_z \partial_{\bar{z}} q^{\mu}(z,\bar{z})+r q^{\mu}(z,\bar{z}), \quad q^{\mu}\equiv \frac{1}{\sqrt{2}}(1+z \bar{z}, z+\bar{z},-i(z-\bar{z}),1-z\bar{z})
\end{eqnarray}
After the redefinition, the Minkowski metric at $\mathscr{I}^+$ is
\begin{eqnarray}
ds^2=- 2 du dr + 2 r^2 dz d\bar{z}
\end{eqnarray}
We notice that the metric component $g_{uu}=0$ by this redefinition. Similarly, at past null infinity, one can redefine the coordinate and the metric in advanced coordinates
\begin{eqnarray}
ds^2= 2 dv dr'+ 2 r'^2 dz' d\bar{z}'.
\end{eqnarray}
At finite $r=r_0$, 
One can make the following identification between these choices of the coordinate system
\begin{eqnarray}
v=u, \quad r'=-r, \quad z'=z
\end{eqnarray}
The Carrollian nature is more transparent in these flat boundary coordinate systems.  At null infinity, we have $dr=0$ and then the metric is
\begin{eqnarray}
ds^2= 0 du dr + 2 r^2 dz d\bar{z}
\end{eqnarray}

This is the degenerate metric at $\mathscr{I}^+$. The Carrollian vector is $n^a \partial_a= \partial_u$. After doing the coordinate transformation, the metric on the boundary is a flat metric $ds^2_{bdry}= 2 dz d\bar{z}$. This completes the analysis for Minkowski spacetime.\\

\subsection{Stress tensor for flat boundary metric}
\label{stress eval}

Now, for any asymptotically flat spacetime that has parameters like $M, N_A, C_{AB}, N_{AB}, D_{AB}$, etc, we will take the flat boundary representative instead of the sphere (see Compere et al \cite{Compere:2016hzt,Compere:2018ylh,Barnich:2016lyg,Donnay:2021wrk,Donnay:2022wvx} for more details). 
\begin{eqnarray}
\label{flat bdry metric}
ds^2&&= \Bigg(\frac{2 M}{r}+O(r^{-2})\Bigg) du^2- 2 \left(1- \frac
{1}{16 r^2}{C^A}_B {C^B}_A+O(r^{-3})\right) du dr\nonumber\\
&&+\Big(r^2 q_{AB} +r C_{AB} +\frac{1}{4}q_{AB} C^C_DC^D_C\Big)dx^A dx^B\nonumber\\
&& +2\Bigg(\frac{1}{2} \partial^B C_{AB} +\frac{2}{3 r} (N_A+\frac{1}{4} C_{AB} \partial_C C^{CB})+O(r^{-2})\Bigg) du dx^A
\end{eqnarray}
where the $q_{AB} dx^A dx^B= 2 dz d\bar{z}$.\\

These parameters satisfy the following constraint equations. These constraint equations are components of the Einstein tensor $E_{uu}=0$ and $E_{uA}=0$.
\begin{eqnarray}
&&\partial_u M=- \frac{1}{8} N^{AB}N_{AB}+\frac{1}{4} \partial_A \partial_B N^{AB}- 4 \pi G \lim_{r \rightarrow \infty}[r^2 T_{uu}^{\mathrm{matter}}]\nonumber
\end{eqnarray}
\begin{eqnarray}
   && \partial_u N_z=  \frac{1}{4} \partial_z(D_z^2 C^{zz}-D_{\bar{z}}^2 C^{\bar{z}\bar{z}})- u \partial_u \partial_z M- T_{uz}\nonumber
\end{eqnarray}
where
\begin{eqnarray}
    T_{uz}= -\frac{1}{4} \partial_z (C_{zz} N^{zz})-\frac{1}{2} C_{zz} D_z N^{zz}+8 \pi G \lim_{r \rightarrow \infty}[r^2 T_{uz}^{\mathrm{matter}}]\nonumber
\end{eqnarray}

The first and second equations can be interpreted as the decay of Bondi mass and angular momentum aspects in the Bondi news (which is the gravitational radiation).\\

The boundary term inaffinity on these solutions can be explicitly obtained for the choice of $n^a,k_a$ written below. 
\begin{eqnarray}
\kappa= \frac{M(u,z,\bar{z})+\frac{1}{8}\partial_u(C_{zz}C_{\bar{z}\bar{z}})
}{r^2}
\end{eqnarray}
Then the on-shell gravitational action is
\begin{eqnarray}
S_{\mathrm{onshell}} = -\frac{1}{8 \pi G_N} \int du\, dz\, d\bar{z}\, r^2  \Bigg( \frac{M(u,z,\bar{z})+\frac{1}{8}\partial_u(C_{zz}C_{\bar{z}\bar{z}})
}{r^2}\Bigg)
\end{eqnarray}
The on-shell action is the \textit{dynamical} mass of the system.\\

The null rigged structure $(\textbf{n},k)$ that decomposes the metric into horizontal and tangential parts are
\begin{eqnarray}
&& k^a=\{-1+O(\frac{1}{r^2}),0,0,0\}\nonumber\\
&& k_a=\{0,1+O(\frac{1}{r^2}),0,0\}, \quad k.k=0\nonumber\\
&& n_a=\{1-\frac
{1}{8 r^2} C_{zz}C_{\bar{z}\bar{z}},-\frac{ M}{r},- \frac{1}{2}\partial_{\bar{z}}C_{zz}- \frac{1}{4r}\partial_{z} C_{\bar{z}\bar{z}}C_{zz},- \frac{1}{2}\partial_{z}C_{\bar{z}\bar{z}}- \frac{1}{4r} C_{\bar{z}\bar{z}}\partial_{\bar{z}}C_{zz}\}\nonumber\\
&& n^a= g^{ab} n_b
\end{eqnarray}
The form of $n_a$ is determined by the following decomposition \footnote{It is important to note that the explicit $1/r^2$ term in $n_a$ is what makes the Brown-York charges consistent with the ones from Wald-Zoupas prescription. Also, compared this metric decomposition with \eqref{withOmega} and note that since we are on the null inifity $\Omega = 0$  }
{\abovedisplayskip=10pt \belowdisplayskip=10pt
$$g_{ab}= q_{ab}-n_a k_b-n_b k_a$$}


The exact form of $n^a$ is a little complicated to write down, but one can expand it in $\frac{1}{r}$. 
 Using the projection operator ${\Pi_a}^b={\delta_a}^b+n_a k^b$, the projection of the normal vector along the null surface defines the Carrollian vector.
\begin{eqnarray}
&&l^a= {\Pi_b}^a n^b, \quad {\Pi_b}^a n_a=0, \quad l^a=n^a+(n.n) k^a, \quad l.l=-n.n\nonumber
\end{eqnarray}

The lower components $l_a$ are obtained using the bulk metric $l_a= g_{ab}l^a$. The decomposition of the inverse metric in terms of horizontal and vertical parts can be written as
{\abovedisplayskip=1pt \belowdisplayskip=1pt
$$g^{ab}=q^{ab}-n^a k^b-k^a n^b$$}
We can also define a projector that has mixed indices, namely, bulk indices and null hypersurface indices.
\begin{eqnarray}
&& {\Pi_i}^a={\delta_i}^a+n_i\, k^a , \qquad   {\Pi_a}^i={\delta_a}^i+ n_a\,  k^i \nonumber\\
&& a=1,2,3,4 \,(\, \text{bulk index}),\quad  i=2,3,4 \,(\,\text{boundary indices})  
\end{eqnarray}
With these projectors, we can write down the Weingarten tensor on the null boundary as 
\begin{eqnarray}
{W_j}^i={\Pi_j}^c\nabla_c n^d {\Pi_d}^i, \quad W= {W_i}^i
\end{eqnarray}
The stress tensor can be computed as 
\begin{eqnarray}
{T_u}^u&&=\frac{-1}{8 \pi G}\left(\frac{2 M}{ r^2}\right)\nonumber\\
{T_u}^z&&=\frac{-1}{8 \pi G}\left(\frac{N_z +\frac{1}{16} \partial_{z}C_{\bar{z}\bar{z}} C_{zz}+\frac{1}{16} C_{\bar{z}\bar{z}} \partial_{z}C_{zz}} { r^2}\right)\nonumber\\
{T_{\bar{z}}}^u&&=\frac{-1}{8 \pi G}\left(\frac{N_{\bar{z}} +\frac{1}{16}C_{\bar{z}\bar{z}} \partial_{\bar{z}}C_{zz}+\frac{1}{16}\partial_{\bar{z}}C_{\bar{z}\bar{z}} C_{zz}} { r^2}\right)\nonumber\\
{T_u}^z&&=\frac{-1}{8 \pi G}\left(-\frac{\partial_{z}N_{\bar{z}\bar{z}}}{2 r^2}\right)\nonumber\\
{T_z}^z&&=\frac{-1}{8 \pi G}\left(\frac{-3N_{\bar{z}\bar{z}} C_{zz}+C_{\bar{z}\bar{z}} N_{zz}+2\partial^{2}_{\bar{z}}C_{zz}-2\partial^{2}_{z}C_{\bar{z}\bar{z}}}{8 r^2}\right)\nonumber\\
{T_{\bar{z}}}^{z}&&=\frac{-1}{8 \pi G}\left(-\frac{ N_{\bar{z}\bar{z}}}{2r}\right)\nonumber\\
{T_{u}}^{\bar{z}}&&=\frac{-1}{8 \pi G}\left(-\frac{\partial_{\bar{z}}N_{zz}}{2 r^2}\right)\nonumber\\
{T_{z}}^{\bar{z}}&&=\frac{-1}{8 \pi G}\left(-\frac{ N_{zz}}{2r}\right)\nonumber\\
{T_{\bar{z}}}^{\bar{z}}&&=\frac{-1}{8 \pi G}\left(\frac{-3C_{\bar{z}\bar{z}} N_{zz}+N_{\bar{z}\bar{z}} C_{zz}-2\partial^{2}_{\bar{z}}C_{zz}+2\partial^{2}_{z}C_{\bar{z}\bar{z}}}{8 r^2}\right)\nonumber\\
\label{stresstensorflat}
\end{eqnarray}
The stress tensor components fall off like $1/r^2$ except for two components ${T_{z}}^{\bar{z}},{T_{\bar{z}}}^{z}$ which have $\frac{1}{r}$ dependence. These would be relevant for understanding the stress tensor due to gravitons in flat spacetime. The ${T_{i}}^u$ where $i=u,z,\bar{z}$ will be used for computing the Brown-York conserved charges (see section \ref{brown}).

\subsubsection*{Trace of stress tensor:}
The trace of the stress tensor can be written as
\begin{eqnarray}
{T_i}^i= -\frac{1}{8 \pi G}\Bigg(\frac{2 M- \frac{1}{4}\partial_u(C_{\bar{z}\bar{z}} C_{zz})}{r^2}\Bigg)+O(1/r^3)
\end{eqnarray}
The stress tensor is not traceless. Its relation to any anomaly of the boundary theory is not clear to us.

\section{BMS Symmetries and charges}
\label{brown}
In this section, we will find the Brown-York charges corresponding to the symmetries of the asymptotic form of the solution. The metric, when written in Bondi gauge, has supertranslation and superrotation symmetries. The vector field generating these symmetries can be written as \cite{Bondi:1962px,Newman:1966ub,Barnich:2011mi,Madler:2016xju,Strominger:2017zoo}
\begin{eqnarray}
\label{symmetry}
\xi= &&Y^A \partial_A+\frac{1}{2} D_A Y^A(u \partial_u-r \partial_r)+ \mathcal{T} \partial_u+\frac{1}{2} D^AD_A \mathcal{T} \partial_r-\frac{1}{r} D^A(\mathcal{T}+\frac{u}{2} D_B Y^B) \partial_A\nonumber\\
&&+\frac{1}{4} D_A D^A D_B Y^B u \partial_r
\end{eqnarray}
The above vector field \footnote{It is crucial to keep the $1/r$ terms in the $\xi^{A}$ part of the Killing vector field. Then only one can show that this is a Killing (conformal) vector field.} generates symmetries at asymptotic infinity in the $1/r$ expansion. The parameters $\mathcal{T}(z,\bar{z})$ are called the supertranslation parameter while $Y^A=\{\mathcal{Y}(z),\bar{\mathcal{Y}}(\bar{z})\}$ are the superrotation parameters. One can understand these symmetries as the conformal symmetries of the Carrollian structure at null infinity. \\

To understand the conformal symmetries for the Carrollian structure at the null boundary, let's start from the Minkowski spacetime metric. The Carrollian structure is described by a degenerate metric $q_{ab}$ and vector field $n^a$ in the kernel of $q_{ab}$. Here the index $x^a= (u,z,\bar{z})$. The degenerate line element can be written as 
\begin{eqnarray}
q_{ab}dx^a dx^b= 0 du^2+ 2 dz d \bar{z}, \quad n^a \partial_a= \partial_u
\end{eqnarray}
The conformal Carrollian symmetries are generated by the vector field
\begin{eqnarray}
\xi_c= (\mathcal{T}+u \alpha) \partial_u +\mathcal{Y} \partial +\bar{\mathcal{Y}} \bar{\partial}, \quad \alpha=\frac{1}{2}(\partial \mathcal{Y}+\bar{\partial} \bar{\mathcal{Y}})
\end{eqnarray}
which satisfies the Carrollian conformal equation
\begin{eqnarray}
\mathcal{L}_{\xi} q_{ab}= 2 \alpha q_{ab}, \quad \mathcal{L}_{\xi} n^a= - \alpha n^a.
\end{eqnarray}
This conformal Carrollian vector $\xi_c$ is the restriction of the vector field of \eqref{symmetry} to the null infinity in the leading order, and taking the boundary metric to be the complex plane. \\

The Quasilocal charges for the supertranslation and superrotations symmetries can be calculated by the Brown-York method using the stress tensor.  
\begin{eqnarray}
\label{QBY}
Q_{BY}&&= -\int_S {T_j}^i\, \xi^j \, \eta_i= -\int_{S} \mu\,\,  ({T_u}^u \xi^u +{T_z}^u \xi^z+{T_{\bar{z}}}^{u} \xi^{\bar{z}}) \nonumber\\
&&=\frac{1}{8 \pi G}\int r^2  dz d\bar{z}\Bigg[\left(\frac{2 M}{r^2}\right)({\mathcal{T}+u \alpha})  \nonumber\\
&& +\left(\frac{N_z +\frac{1}{16} \partial_{z}C_{\bar{z}\bar{z}} C_{zz}+\frac{1}{16} C_{\bar{z}\bar{z}} \partial_{z}C_{zz}} { r^2}\right)\mathcal{Y}+\left(\frac{N_{\bar{z}} +\frac{1}{16}C_{\bar{z}\bar{z}} \partial_{\bar{z}}C_{zz}+\frac{1}{16}\partial_{\bar{z}}C_{\bar{z}\bar{z}} C_{zz}} { r^2}\right) \bar{\mathcal{Y}}\Bigg]+\cdots\nonumber\\
\end{eqnarray}
We have used the volume element at null infinity $\eta_i= k_i \, \mu  $ where $k_u=1, k_{z}=k_{\bar{z}}=0$ and the spatial volume element $\mu= r^2-\frac{1}{2} C_{zz} C_{\bar{z}\bar{z}} $. The dotted terms are of the order $O(\frac{1}{r})$ and will vanish in the strict $r\rightarrow \infty$ limit. \\

The charges (Wald-Zoupas) for BMS symmetries have been calculated using the covariant phase space method \cite{Barnich:2011mi}. The charges are one form on the solution space and can be written as
\begin{eqnarray}
    \not\delta \mathcal{Q}_{\xi}[\delta \mathcal{X},\mathcal{X}]=\delta (Q_s[\mathcal{X}])+\Theta_s[\delta \mathcal{X},\mathcal{X}]
\end{eqnarray}
where the integrable charge one form can be written as
\begin{eqnarray}
Q_s[\mathcal{X}]=\frac{1}{16 \pi G} \int d^2 z [4 \mathcal{T} M]+Y^A(2 N_A+\frac{1}{16} \partial_A(C^{BC}C_{BC})).
\end{eqnarray}
The non-integrable comes from the news tensor and can be written as
\begin{eqnarray}
 \Theta_s[\delta \mathcal{X},\mathcal{X}]=\frac{1}{16 \pi G} \int d^2 z[\frac{\mathcal{T}}{2} N_{AB}\delta C^{AB}].   
\end{eqnarray}
Hence, our answer for the charges \eqref{QBY} matches with Barnich et. al. \cite{Barnich:2011mi} up to the non-integrable charges. 
The Brown York charges depend only on the solution space $\mathcal{X}=\{M, N_A,C_{AB},N_{AB}\}$, not on its variation. Hence, it is reasonable to argue that the Brown York charges only give the integrable part of Wald Zoupas' charges.


\subsubsection*{Discussion of corner terms:} In this whole formalism, there are corner terms that we have ignored, and there are choices of boundary terms (on the null boundary) that can change the corner terms. One of the choices is $\frac{1}{8 \pi G} \int_{\mathcal{N}} (\Theta+\kappa)$. Here $\Theta$ is the expansion. These choices change the corner terms. These corner terms are integrated over codimension 2 surfaces.
\begin{eqnarray}
&&\Theta= \left(\frac{-2 M}{ r^2}\right)\\
&& \kappa=\left(\frac{8 M+N_{\bar{z}\bar{z}} C_{zz}+C_{\bar{z}\bar{z}} N_{zz}}{8 r^2}\right)
\end{eqnarray}

\subsection{The Kerr black hole: A case study}
\label{the Kerr bh}

The Kerr metric, which describes a rotating black hole in Boyer-Lindquist coordinates, is given by:
\begin{equation}
    ds^2 = -\left( 1 - \frac{2 M r}{\Sigma} \right) dt^2 - \frac{4 M a r\sin^2\theta}{\Sigma} dt\, d\phi + \frac{\Sigma}{\Delta} dr^2 + \Sigma\, d\theta^2 + \left( r^2 + a^2 + \frac{2 M a^2r\sin^2\theta}{\Sigma} \right) \sin^2\theta\, d\phi^2,
\end{equation}
where the functions \(\Delta\) and \(\Sigma\) are defined as:
\begin{align}
    \Delta &= r^2 - 2 M r + a^2, \\
    \Sigma &= r^2 + a^2\cos^2\theta.
\end{align}
Here, \(M \) is the mass of the black hole, and \(a\) is the specific angular momentum per unit mass.\\
Now we write the metric in BMS gauge following \cite{SJFletcher:2003,Barnich:2011mi}, This gives the various metric components as
\begin{eqnarray}
    &&g_{uu}=-\frac{a \, M \cos (2 \theta ) \csc (\theta )}{r^2}+\frac{2 M}{r}-1+O(1/r^3)\\
    &&g_{ur}=-\frac{a^2 \cos (2 \theta )}{4 r^2}-1+O(1/r^3)\\
    &&g_{u\phi}=-\frac{2 a \, M \sin ^2(\theta )}{r}+O(1/r^2)\\
&&g_{\theta\theta}=r^2+\frac{a}{\mathrm{sin}\, \theta} r+\frac{a^2}{2\, \mathrm{sin}^2\, \theta} +O(1/r)\\
   &&g_{\phi\phi}=r^2 \mathrm{sin}^2\, \theta-a \mathrm{sin}\, \theta  \, r+\frac{a^2}{2}+O(1/r)\\
  && g_{\theta\phi}=O(1/r)\\
  &&g_{u\theta}=\frac{a}{2}\frac{\mathrm{ cos}\, \theta}{ \mathrm{sin}^2\, \theta}+\frac{a\, \mathrm{cos}\,\theta}{4} (8 M+\frac{a}{\mathrm{sin}^3\,\theta})r^{-1}+O(1/r^2).
\end{eqnarray}
Now the parameters of the asymptotically flat spacetime can be read off as
\begin{eqnarray}
&&m_{B}(u,\theta,\phi)=M, \quad N_{AB}=0\\
 && C_{\theta\theta}= \frac{a}{\mathrm{sin}\, \theta}, \quad C_{\phi\phi}=-a\, \mathrm{sin}\, \theta, \quad C_{\theta\phi}=0\\
 && N_{\theta}=3 M \, a \mathrm{cos}\, \theta+\frac{a^2}{8} \frac{\mathrm{cos}\, \theta}{\mathrm{sin}^3\,\theta}, \quad N_{\phi}=-3 a \, M \mathrm{sin}^2\theta
\end{eqnarray}
The normal and auxiliary vector for the Kerr spacetime ($x^a=\{r,u,\theta,\phi\}$) can be written as
\begin{eqnarray}
   && k_a=\{0,1,0,0\}, \quad k^a=g^{ab}k_b\\
    && n_a=\{1,1-\frac{M}{r}, -\frac{1}{2} a \frac{\cot\,\theta}{\sin\,\theta} ,0\}
\end{eqnarray}
These vectors are fixed by the following decomposition of the metric \footnote{Here the extra term of $k_a k_b\, n.n$ is needed because the $n.n=-1+O(1/r^2)$. With that, the metric can be decomposed in terms of horizontal, vertical, and transverse subspaces. In previous cases of asymptotically flat spacetime, the norm of the normal vector is $O(1/r^3)$; hence, it was not needed where the metric goes most $O(1/r^2)$. To avoid this subtlety, one can do the coordinate transformation like the one done in \eqref{weyl mink}. This will get rid of $-1$ in the $g_{uu}$ term. And then we don't need to subtract the extra $k_a k_b n.n$ term.}
\begin{eqnarray}
    g_{ab}=q_{ab}-k_a n_b-n_bk_a-k_a k_b n.n
\end{eqnarray}
With these normal vectors, we can compute the Weingarten tensor and then the stress tensor. The stress tensor can be computed as
\begin{eqnarray}
 {T_u}^u&&=\frac{-1}{8 \pi G}\left(\frac{2 M}{r^2}\right)+O(1/r^3)\nonumber\\
{T_{z}}^u&&=\frac{-1}{8 \pi G}\left(\frac{a \sin (\theta ) \left(3 \cot (\theta ) \left(a \csc ^3(\theta )+8 M\right)-4 a \cos (\theta )\right)}{8 r^2}\right)+O(1/r^3)\nonumber\\
{T_{\bar{z}}}^u&&=\frac{-1}{8 \pi G}\left(-\frac{3 J M \sin ^2(\theta )}{r^2}\right)+O(1/r^3)\nonumber\\
\end{eqnarray}
All other components are $O(1/r^3)$, and are not relevant for the computation of charges or boundary stress tensor.\\
\subsection*{Brown-York charges for the Kerr black hole}
The vector field generating the symmetries for the Kerr black hole is
\begin{eqnarray}
    \xi\equiv \xi^a \partial_a= \mathcal{T}(\theta,\phi)\partial_u+Y^{\phi}(\theta,\phi)\partial_{\phi}.
\end{eqnarray}
Hence, the Brown-York charges are
\begin{eqnarray}
    Q_{BY}&&= -\int_S T^i_j \xi^j \eta_i \nonumber\\
&&=\frac{1}{8 \pi G}\int d \theta d \phi\,   r^2  \sin \theta \Bigg[\frac{2 M}{ r^2}\mathcal{T}+Y^{\phi} (-\frac{3 a \, M \sin ^2(\theta )}{r^2})\Bigg]+\cdots\nonumber\\
\end{eqnarray}
We have used the volume element at null infinity $\eta_i= k_i \, \mu \,\,, k_u=1,k_{\theta}=k_{\phi}=0 $ and the spatial volume element $\mu= r^2+O(r)$. The $\cdots$ in the above equation are $O(1/r^3)$ and won't contribute to the charges. The global charges when $\mathcal{T}$ and $Y^{\phi}$ are independent of angular coordinates, then the Brown-York charges gives the familiar mass $\frac{M}{G}$ and angular momentum $J= -\frac{a \, M}{G}$ of the Kerr black holes. \footnote{The overall sign for the charges are consistent with \cite{Barnich:2011mi,Wald:1999wa}. } \\

We can decompose the Killing vectors in spherical harmonics and find charges for each mode.
\begin{eqnarray}
    \mathcal{T}(\theta,\phi)= \sum_{l,m} Y_{l,m} (\theta, \phi)\mathcal{T}_{lm}\quad 
\end{eqnarray}
Only $l=m=0$ modes of the supertranslation will give the non-vanishing charge because the integral of spherical harmonics over the whole sphere is zero except for $l=m=0$ mode. For superotations, it is more convenient to transform to the complex coordinates ($z,\bar{z}$).
\begin{eqnarray}
    z= e^{i \phi} \cot\, \theta/2, \quad Y^{\phi}=\frac{-i}{2 z} Y^z(z)+\frac{i}{2 \bar{z}} Y^{\bar{z}}(\bar{z})
\end{eqnarray}
Then we can decompose these vector fields as
\begin{eqnarray}
    Y^z=\sum_n a_n  L_n, \quad   Y^{\bar{z}}=\sum_m a_m  \bar{L}_m,\quad  L_n=-z^{n+1} \partial_z, \quad \bar{L}_m=-\bar{z}^{m+1}\partial_{\bar{z}}
\end{eqnarray}
Then we can compute the charges for each mode as \footnote{The easiest way to do the following integrals is to transform them into angular variables $(\theta,\phi)$ using the stereographic map.}
\begin{eqnarray}
    Q_{L_n}= -\frac{3 a M}{8 \pi G}\int  \frac{2\, dz\, d\bar{z}}{i (1+z\bar{z})^4} 4 (z \bar{z})^2 (\frac{-i}{2 z}) z^{n+1}= -\delta^n_0\frac{i a \, M }{2 G}\\
     Q_{\bar{L}_n}= -\frac{3 a M}{8 \pi G}\int  \frac{2\, dz\, d\bar{z}}{i (1+z\bar{z})^4} 4 (z \bar{z})^2 (\frac{-i}{2 z}) z^{n+1}= \delta^n_0\frac{i a \, M }{2 G}
\end{eqnarray}
And the global charge for  $\partial_{\phi}= -i (L_0-\bar{L}_0)$ is $Q= -\frac{a\,M}{ G}$. For the central extension of these algebras (see section 4 of \cite{Barnich:2011mi}). 
\section{Discussions:}
In this article, we explicitly calculated the holographic stress tensor \eqref{stresstensorflat} for asymptotically flat spacetime. This is the direct analog of the holographic stress tensor of AdS spacetime of Balasubramanian et al. \cite{Balasubramanian:1999re}. Following a similar strategy to AdS spacetime, we can study anomalies and holographic renormalization in Celestial/Carrollian holography. We also calculated the Brown-York charges for the symmetries, and they can be compared with the Wald-Zoupas charges. There are several future directions one can work on. \begin{itemize}
    \item We need to understand the contribution to the stress tensor from the joints $i^0$.
    \item We need to understand how the holographic stress tensor \eqref{stresstensorflat} acts on other fields/operators of the theory. Then we can compare it with the stress tensor of \cite{Kapec:2016jld,Kapec:2017gsg} for the gravitons.
    \item The contribution of soft charges from the time like infinity $i^{\pm}$ needs to be understood. We also need to understand boundary action $\int  du \,d^2z\, \kappa=\int du\, d^2 z\left(\frac{8 M+N_{\bar{z}\bar{z}} C_{zz}+C_{\bar{z}\bar{z}} N_{zz}}{8 }\right)$ of ours to the  onshell action of He et.al. \cite{He:2024vlp}.
\end{itemize}

\begin{acknowledgments}
The work of HK is partly supported by NSF grants PHY-2210533 and PHY-2210562.  J.B. is supported by the U.S. Department of Energy, Office of Science, Office of Nuclear Physics, grant No. DE-FG-02-08ER41450.
\end{acknowledgments}
 
\appendix

\section{A gentle introduction to structure at null hypersurface}
\label{geo null}
In this appendix, we reviewed the geometrical structure at the null surfaces, borrowing intuition from the spacelike or timelike surfaces. We consider a 4-dimensional spacetime $\mathscr{M}$ with Lorentzian signature metric $g_{ab}(x)$. In this spacetime, a 3-dimensional hypersurface $\Sigma$ is defined via $\phi(x^a) = 0$. A normal form to such a surface is $s_{a} = \frac{\pm\partial_{a}\phi}{\sqrt{g^{ab}\partial_{a}\phi \partial_{b}\phi}}$. The $\pm$ sign depends on whether $\Sigma$ is timelike or spacelike, respectively. Then the normal vector is obtained using the inverse metric, $s^a = g^{ab}s_{b}$. This definition of the normal form breaks down for null surfaces as the normalization $g^{ab}\partial_{a}\phi \partial_{b}\phi$ becomes zero.\\
Instead, for the null surface, we define
\begin{eqnarray}
n_{a}=-\partial_a \phi, \quad n^a= g^{ab} n_a, \quad n_a n^a=0
\end{eqnarray}
The sign is chosen so that $n^a$ is future-directed when $\phi$ increases to the future.
Let's evaluate
\begin{eqnarray}
n^{b}\nabla_{b}n^a &&= n^b\nabla_{b} (g^{ab}\partial_{b}\phi)\\
&&= \frac{1}{2} g^{ab}\nabla_{b}(n^c n_c)
\end{eqnarray}
$$ $$ where $\nabla_{a}$ is the Levi-Civita connection. Note that $n^cn_c = 0$ everywhere on $\Sigma$; hence its gradient will be non-zero only away from the hypersurface, i.e., in the direction of the normal hence, the gradient of $n^cn_c $ must be proportional to $n^a$,
\begin{eqnarray}
\label{generalzied_geodesic}
n^{b}\nabla_{b}n^a = \kappa n^a
\end{eqnarray}
where $\kappa(x)$ is some scalar function called inaffinity. Note that the normal vector field $n^a$ is tangent to $\Sigma$. The normal vector satisfies the generalized geodesic equation \eqref{generalzied_geodesic}. The hypersurface $\Sigma$ is generated by a null geodesic, and $n^a$ is tangent to the geodesic. \\ 

The null geodesics are parametrized parameter $\lambda$ (not always affine). Then the displacement along the generator is $dx^a= n^a d \lambda$. Then on $\Sigma$, a coordinate system is placed that is compatible with generators. Hence, One of the coordinate is taken to be $\lambda$ and two additional coordinates $\theta^A$. 
$$y^i = (\lambda, \theta^A)$$
where A = 1, 2.  The induced metric on the hypersurface can be written as
\begin{eqnarray}
ds^2_\Sigma &&= g_{ab}\frac{dx^a}{d\theta^A}\frac{dx^b}{d\theta^B}d\theta^A d\theta^B\\
&&\equiv \sigma_{AB}d\theta^A d\theta^B, \quad e^a_A= \Big(\frac{\partial x^a}{\partial \theta^A}\Big)_{\lambda}
\end{eqnarray}
Such an induced metric is degenerate. What this means is that one of the directions is ``missing" on the null hypersurface, or more precisely we cannot correctly distinguish between the tangent and normal (inherent degeneracy on null hypersurfaces). To solve this, an extra structure in the form of an auxiliary vector $k^a$ satisfying $k^a n_a = -1$ and $k^a e^A_a = 0$, with the help of which we decompose the inverse spacetime metric as,
\begin{eqnarray}
g^{ab} = -n^a k^b - k^a n^b + \sigma^{AB}e^a_A e^b_B
\end{eqnarray}
where $\sigma^{AB}$ is the inverse of $\sigma_{AB}$. Thus $k$ vector field provides the ``missing" direction. Such null hypersurfaces with the above extra geometric structure are what adequately define null infinity. (see Poisson's book \cite{Poisson_2004}- chapter 3 for an example and more details). The null rigged structure's $k$ vector field will be identified with this auxiliary $k$ vector field.

\bibliographystyle{JHEP}
\bibliography{ref}

\providecommand{\href}[2]{#2}\begingroup\raggedright\begin{thebibliography}{10}

\bibitem{Raju:2019qjq}
S.~Raju, \emph{{Is Holography Implicit in Canonical Gravity?}}, \href{https://doi.org/10.1142/S0218271819440115}{\emph{Int. J. Mod. Phys. D} {\bfseries 28} (2019) 1944011} [\href{https://arxiv.org/abs/1903.11073}{{\ttfamily 1903.11073}}].

\bibitem{Witten:1998qj}
E.~Witten, \emph{{Anti-de Sitter space and holography}}, \href{https://doi.org/10.4310/ATMP.1998.v2.n2.a2}{\emph{Adv. Theor. Math. Phys.} {\bfseries 2} (1998) 253} [\href{https://arxiv.org/abs/hep-th/9802150}{{\ttfamily hep-th/9802150}}].

\bibitem{Aharony:1999ti}
O.~Aharony, S.S.~Gubser, J.M.~Maldacena, H.~Ooguri and Y.~Oz, \emph{{Large N field theories, string theory and gravity}}, \href{https://doi.org/10.1016/S0370-1573(99)00083-6}{\emph{Phys. Rept.} {\bfseries 323} (2000) 183} [\href{https://arxiv.org/abs/hep-th/9905111}{{\ttfamily hep-th/9905111}}].

\bibitem{Maldacena:1997re}
J.M.~Maldacena, \emph{{The Large N limit of superconformal field theories and supergravity}}, \href{https://doi.org/10.4310/ATMP.1998.v2.n2.a1}{\emph{Adv. Theor. Math. Phys.} {\bfseries 2} (1998) 231} [\href{https://arxiv.org/abs/hep-th/9711200}{{\ttfamily hep-th/9711200}}].

\bibitem{Balasubramanian:1999re}
V.~Balasubramanian and P.~Kraus, \emph{{A Stress tensor for Anti-de Sitter gravity}}, \href{https://doi.org/10.1007/s002200050764}{\emph{Commun. Math. Phys.} {\bfseries 208} (1999) 413} [\href{https://arxiv.org/abs/hep-th/9902121}{{\ttfamily hep-th/9902121}}].

\bibitem{Brown:1992br}
J.D.~Brown and J.W.~York, Jr., \emph{{Quasilocal energy and conserved charges derived from the gravitational action}}, \href{https://doi.org/10.1103/PhysRevD.47.1407}{\emph{Phys. Rev. D} {\bfseries 47} (1993) 1407} [\href{https://arxiv.org/abs/gr-qc/9209012}{{\ttfamily gr-qc/9209012}}].

\bibitem{Bianchi:2001kw}
M.~Bianchi, D.Z.~Freedman and K.~Skenderis, \emph{{Holographic renormalization}}, \href{https://doi.org/10.1016/S0550-3213(02)00179-7}{\emph{Nucl. Phys. B} {\bfseries 631} (2002) 159} [\href{https://arxiv.org/abs/hep-th/0112119}{{\ttfamily hep-th/0112119}}].

\bibitem{Skenderis:2002wp}
K.~Skenderis, \emph{{Lecture notes on holographic renormalization}}, \href{https://doi.org/10.1088/0264-9381/19/22/306}{\emph{Class. Quant. Grav.} {\bfseries 19} (2002) 5849} [\href{https://arxiv.org/abs/hep-th/0209067}{{\ttfamily hep-th/0209067}}].

\bibitem{Henningson:1998gx}
M.~Henningson and K.~Skenderis, \emph{{The Holographic Weyl anomaly}}, \href{https://doi.org/10.1088/1126-6708/1998/07/023}{\emph{JHEP} {\bfseries 07} (1998) 023} [\href{https://arxiv.org/abs/hep-th/9806087}{{\ttfamily hep-th/9806087}}].

\bibitem{deHaro:2000vlm}
S.~de~Haro, S.N.~Solodukhin and K.~Skenderis, \emph{{Holographic reconstruction of space-time and renormalization in the AdS / CFT correspondence}}, \href{https://doi.org/10.1007/s002200100381}{\emph{Commun. Math. Phys.} {\bfseries 217} (2001) 595} [\href{https://arxiv.org/abs/hep-th/0002230}{{\ttfamily hep-th/0002230}}].

\bibitem{Brown:1986nw}
J.D.~Brown and M.~Henneaux, \emph{{Central Charges in the Canonical Realization of Asymptotic Symmetries: An Example from Three-Dimensional Gravity}}, \href{https://doi.org/10.1007/BF01211590}{\emph{Commun. Math. Phys.} {\bfseries 104} (1986) 207}.

\bibitem{Myers:2010tj}
R.C.~Myers and A.~Sinha, \emph{{Holographic c-theorems in arbitrary dimensions}}, \href{https://doi.org/10.1007/JHEP01(2011)125}{\emph{JHEP} {\bfseries 01} (2011) 125} [\href{https://arxiv.org/abs/1011.5819}{{\ttfamily 1011.5819}}].

\bibitem{Myers:2010xs}
R.C.~Myers and A.~Sinha, \emph{{Seeing a c-theorem with holography}}, \href{https://doi.org/10.1103/PhysRevD.82.046006}{\emph{Phys. Rev. D} {\bfseries 82} (2010) 046006} [\href{https://arxiv.org/abs/1006.1263}{{\ttfamily 1006.1263}}].

\bibitem{Komargodski:2011vj}
Z.~Komargodski and A.~Schwimmer, \emph{{On Renormalization Group Flows in Four Dimensions}}, \href{https://doi.org/10.1007/JHEP12(2011)099}{\emph{JHEP} {\bfseries 12} (2011) 099} [\href{https://arxiv.org/abs/1107.3987}{{\ttfamily 1107.3987}}].

\bibitem{Karateev:2023mrb}
D.~Karateev, Z.~Komargodski, J.a.~Penedones and B.~Sahoo, \emph{{Trace anomalies and the graviton-dilaton amplitude}}, \href{https://doi.org/10.1007/JHEP11(2024)067}{\emph{JHEP} {\bfseries 11} (2024) 067} [\href{https://arxiv.org/abs/2312.09308}{{\ttfamily 2312.09308}}].

\bibitem{Calabrese:2004eu}
P.~Calabrese and J.L.~Cardy, \emph{{Entanglement entropy and quantum field theory}}, \href{https://doi.org/10.1088/1742-5468/2004/06/P06002}{\emph{J. Stat. Mech.} {\bfseries 0406} (2004) P06002} [\href{https://arxiv.org/abs/hep-th/0405152}{{\ttfamily hep-th/0405152}}].

\bibitem{Ryu:2006ef}
S.~Ryu and T.~Takayanagi, \emph{{Aspects of Holographic Entanglement Entropy}}, \href{https://doi.org/10.1088/1126-6708/2006/08/045}{\emph{JHEP} {\bfseries 08} (2006) 045} [\href{https://arxiv.org/abs/hep-th/0605073}{{\ttfamily hep-th/0605073}}].

\bibitem{Strominger:2017zoo}
A.~Strominger, \emph{{Lectures on the Infrared Structure of Gravity and Gauge Theory}} (3, 2017), [\href{https://arxiv.org/abs/1703.05448}{{\ttfamily 1703.05448}}].

\bibitem{Ashtekar:2023wfn}
A.~Ashtekar and N.~Khera, \emph{{Unified treatment of null and spatial infinity III: asymptotically minkowski space-times}}, \href{https://doi.org/10.1007/JHEP02(2024)210}{\emph{JHEP} {\bfseries 02} (2024) 210} [\href{https://arxiv.org/abs/2311.14130}{{\ttfamily 2311.14130}}].

\bibitem{Ashtekar:2023zul}
A.~Ashtekar and N.~Khera, \emph{{Unified treatment of null and spatial infinity IV: angular momentum at null and spatial infinity}}, \href{https://doi.org/10.1007/JHEP01(2024)085}{\emph{JHEP} {\bfseries 01} (2024) 085} [\href{https://arxiv.org/abs/2311.14190}{{\ttfamily 2311.14190}}].

\bibitem{Parattu:2015gga}
K.~Parattu, S.~Chakraborty, B.R.~Majhi and T.~Padmanabhan, \emph{{A Boundary Term for the Gravitational Action with Null Boundaries}}, \href{https://doi.org/10.1007/s10714-016-2093-7}{\emph{Gen. Rel. Grav.} {\bfseries 48} (2016) 94} [\href{https://arxiv.org/abs/1501.01053}{{\ttfamily 1501.01053}}].

\bibitem{Chandrasekaran:2020wwn}
V.~Chandrasekaran and A.J.~Speranza, \emph{{Anomalies in gravitational charge algebras of null boundaries and black hole entropy}}, \href{https://doi.org/10.1007/JHEP01(2021)137}{\emph{JHEP} {\bfseries 01} (2021) 137} [\href{https://arxiv.org/abs/2009.10739}{{\ttfamily 2009.10739}}].

\bibitem{Chandrasekaran:2021hxc}
V.~Chandrasekaran, E.E.~Flanagan, I.~Shehzad and A.J.~Speranza, \emph{{Brown-York charges at null boundaries}}, \href{https://doi.org/10.1007/JHEP01(2022)029}{\emph{JHEP} {\bfseries 01} (2022) 029} [\href{https://arxiv.org/abs/2109.11567}{{\ttfamily 2109.11567}}].

\bibitem{Aghapour:2018icu}
S.~Aghapour, G.~Jafari and M.~Golshani, \emph{{On variational principle and canonical structure of gravitational theory in double-foliation formalism}}, \href{https://doi.org/10.1088/1361-6382/aaef9e}{\emph{Class. Quant. Grav.} {\bfseries 36} (2019) 015012} [\href{https://arxiv.org/abs/1808.07352}{{\ttfamily 1808.07352}}].

\bibitem{Jafari:2019bpw}
G.~Jafari, \emph{{Stress Tensor on Null Boundaries}}, \href{https://doi.org/10.1103/PhysRevD.99.104035}{\emph{Phys. Rev. D} {\bfseries 99} (2019) 104035} [\href{https://arxiv.org/abs/1901.04054}{{\ttfamily 1901.04054}}].

\bibitem{Kapec:2016jld}
D.~Kapec, P.~Mitra, A.-M.~Raclariu and A.~Strominger, \emph{{2D Stress Tensor for 4D Gravity}}, \href{https://doi.org/10.1103/PhysRevLett.119.121601}{\emph{Phys. Rev. Lett.} {\bfseries 119} (2017) 121601} [\href{https://arxiv.org/abs/1609.00282}{{\ttfamily 1609.00282}}].

\bibitem{Ruzziconi:2024kzo}
R.~Ruzziconi and A.~Saha, \emph{{Holographic Carrollian currents for massless scattering}}, \href{https://doi.org/10.1007/JHEP01(2025)169}{\emph{JHEP} {\bfseries 01} (2025) 169} [\href{https://arxiv.org/abs/2411.04902}{{\ttfamily 2411.04902}}].

\bibitem{Saha:2023hsl}
A.~Saha, \emph{{Carrollian approach to 1 + 3D flat holography}}, \href{https://doi.org/10.1007/JHEP06(2023)051}{\emph{JHEP} {\bfseries 06} (2023) 051} [\href{https://arxiv.org/abs/2304.02696}{{\ttfamily 2304.02696}}].

\bibitem{Bhattacharya:2024fbz}
K.~Bhattacharya, S.~Dey and B.R.~Majhi, \emph{{Gravity and fluid dynamic correspondence on a null hypersurface: inconsistencies and advancement}},  \href{https://arxiv.org/abs/2411.06914}{{\ttfamily 2411.06914}}.

\bibitem{Ciambelli:2023mvj}
L.~Ciambelli and L.~Lehner, \emph{{Fluid-gravity correspondence and causal first-order relativistic viscous hydrodynamics}}, \href{https://doi.org/10.1103/PhysRevD.108.126019}{\emph{Phys. Rev. D} {\bfseries 108} (2023) 126019} [\href{https://arxiv.org/abs/2310.15427}{{\ttfamily 2310.15427}}].

\bibitem{Wald:1999wa}
R.M.~Wald and A.~Zoupas, \emph{{A General definition of 'conserved quantities' in general relativity and other theories of gravity}}, \href{https://doi.org/10.1103/PhysRevD.61.084027}{\emph{Phys. Rev. D} {\bfseries 61} (2000) 084027} [\href{https://arxiv.org/abs/gr-qc/9911095}{{\ttfamily gr-qc/9911095}}].

\bibitem{Barnich:2011mi}
G.~Barnich and C.~Troessaert, \emph{{BMS charge algebra}}, \href{https://doi.org/10.1007/JHEP12(2011)105}{\emph{JHEP} {\bfseries 12} (2011) 105} [\href{https://arxiv.org/abs/1106.0213}{{\ttfamily 1106.0213}}].

\bibitem{Ashtekar:2024bpi}
A.~Ashtekar and S.~Speziale, \emph{{Null infinity as a weakly isolated horizon}}, \href{https://doi.org/10.1103/PhysRevD.110.044048}{\emph{Phys. Rev. D} {\bfseries 110} (2024) 044048} [\href{https://arxiv.org/abs/2402.17977}{{\ttfamily 2402.17977}}].

\bibitem{Ashtekar:2024mme}
A.~Ashtekar and S.~Speziale, \emph{{Horizons and null infinity: A fugue in four voices}}, \href{https://doi.org/10.1103/PhysRevD.109.L061501}{\emph{Phys. Rev. D} {\bfseries 109} (2024) L061501} [\href{https://arxiv.org/abs/2401.15618}{{\ttfamily 2401.15618}}].

\bibitem{Ashtekar:2024stm}
A.~Ashtekar and S.~Speziale, \emph{{Null infinity and horizons: A new approach to fluxes and charges}}, \href{https://doi.org/10.1103/PhysRevD.110.044049}{\emph{Phys. Rev. D} {\bfseries 110} (2024) 044049} [\href{https://arxiv.org/abs/2407.03254}{{\ttfamily 2407.03254}}].

\bibitem{Wald:1984rg}
R.M.~Wald, \emph{{General Relativity}}, Chicago Univ. Pr., Chicago, USA (1984), \href{https://doi.org/10.7208/chicago/9780226870373.001.0001}{10.7208/chicago/9780226870373.001.0001}.

\bibitem{Ciambelli:2023ott}
L.~Ciambelli, A.~Delfante, R.~Ruzziconi and C.~Zwikel, \emph{{Symmetries and charges in Weyl-Fefferman-Graham gauge}}, \href{https://doi.org/10.1103/PhysRevD.108.126003}{\emph{Phys. Rev. D} {\bfseries 108} (2023) 126003} [\href{https://arxiv.org/abs/2308.15480}{{\ttfamily 2308.15480}}].

\bibitem{Geiller:2024amx}
M.~Geiller and C.~Zwikel, \emph{{The partial Bondi gauge: Gauge fixings and asymptotic charges}}, \href{https://doi.org/10.21468/SciPostPhys.16.3.076}{\emph{SciPost Phys.} {\bfseries 16} (2024) 076} [\href{https://arxiv.org/abs/2401.09540}{{\ttfamily 2401.09540}}].

\bibitem{Geiller:2024ryw}
M.~Geiller, A.~Laddha and C.~Zwikel, \emph{{Symmetries of the gravitational scattering in the absence of peeling}}, \href{https://doi.org/10.1007/JHEP12(2024)081}{\emph{JHEP} {\bfseries 12} (2024) 081} [\href{https://arxiv.org/abs/2407.07978}{{\ttfamily 2407.07978}}].

\bibitem{Freidel:2022bai}
L.~Freidel and P.~Jai-akson, \emph{{Carrollian hydrodynamics from symmetries}}, \href{https://doi.org/10.1088/1361-6382/acb194}{\emph{Class. Quant. Grav.} {\bfseries 40} (2023) 055009} [\href{https://arxiv.org/abs/2209.03328}{{\ttfamily 2209.03328}}].

\bibitem{Freidel:2022vjq}
L.~Freidel and P.~Jai-akson, \emph{{Carrollian hydrodynamics and symplectic structure on stretched horizons}},  \href{https://arxiv.org/abs/2211.06415}{{\ttfamily 2211.06415}}.

\bibitem{Freidel:2024emv}
L.~Freidel and P.~Jai-akson, \emph{{Geometry of Carrollian Stretched Horizons}},  \href{https://arxiv.org/abs/2406.06709}{{\ttfamily 2406.06709}}.

\bibitem{Riello:2024uvs}
A.~Riello and L.~Freidel, \emph{{Renormalization of conformal infinity as a stretched horizon}}, \href{https://doi.org/10.1088/1361-6382/ad5cbb}{\emph{Class. Quant. Grav.} {\bfseries 41} (2024) 175013} [\href{https://arxiv.org/abs/2402.03097}{{\ttfamily 2402.03097}}].

\bibitem{Donnay:2019jiz}
L.~Donnay and C.~Marteau, \emph{{Carrollian Physics at the Black Hole Horizon}}, \href{https://doi.org/10.1088/1361-6382/ab2fd5}{\emph{Class. Quant. Grav.} {\bfseries 36} (2019) 165002} [\href{https://arxiv.org/abs/1903.09654}{{\ttfamily 1903.09654}}].

\bibitem{Mars_1993}
M.~Mars and J.M.M.~Senovilla, \emph{Geometry of general hypersurfaces in spacetime: junction conditions}, \href{https://doi.org/10.1088/0264-9381/10/9/026}{\emph{Classical and Quantum Gravity} {\bfseries 10} (1993) 1865–1897}.

\bibitem{Gourgoulhon:2005ng}
E.~Gourgoulhon and J.L.~Jaramillo, \emph{{A 3+1 perspective on null hypersurfaces and isolated horizons}}, \href{https://doi.org/10.1016/j.physrep.2005.10.005}{\emph{Phys. Rept.} {\bfseries 423} (2006) 159} [\href{https://arxiv.org/abs/gr-qc/0503113}{{\ttfamily gr-qc/0503113}}].

\bibitem{Donnay:2022aba}
L.~Donnay, A.~Fiorucci, Y.~Herfray and R.~Ruzziconi, \emph{{Carrollian Perspective on Celestial Holography}}, \href{https://doi.org/10.1103/PhysRevLett.129.071602}{\emph{Phys. Rev. Lett.} {\bfseries 129} (2022) 071602} [\href{https://arxiv.org/abs/2202.04702}{{\ttfamily 2202.04702}}].

\bibitem{Donnay:2022wvx}
L.~Donnay, A.~Fiorucci, Y.~Herfray and R.~Ruzziconi, \emph{{Bridging Carrollian and celestial holography}}, \href{https://doi.org/10.1103/PhysRevD.107.126027}{\emph{Phys. Rev. D} {\bfseries 107} (2023) 126027} [\href{https://arxiv.org/abs/2212.12553}{{\ttfamily 2212.12553}}].

\bibitem{Barnich:2010eb}
G.~Barnich and C.~Troessaert, \emph{{Aspects of the BMS/CFT correspondence}}, \href{https://doi.org/10.1007/JHEP05(2010)062}{\emph{JHEP} {\bfseries 05} (2010) 062} [\href{https://arxiv.org/abs/1001.1541}{{\ttfamily 1001.1541}}].

\bibitem{Compere:2016hzt}
G.~Comp\`ere and J.~Long, \emph{{Classical static final state of collapse with supertranslation memory}}, \href{https://doi.org/10.1088/0264-9381/33/19/195001}{\emph{Class. Quant. Grav.} {\bfseries 33} (2016) 195001} [\href{https://arxiv.org/abs/1602.05197}{{\ttfamily 1602.05197}}].

\bibitem{Compere:2018ylh}
G.~Comp\`ere, A.~Fiorucci and R.~Ruzziconi, \emph{{Superboost transitions, refraction memory and super-Lorentz charge algebra}}, \href{https://doi.org/10.1007/JHEP11(2018)200}{\emph{JHEP} {\bfseries 11} (2018) 200} [\href{https://arxiv.org/abs/1810.00377}{{\ttfamily 1810.00377}}].

\bibitem{Barnich:2016lyg}
G.~Barnich and C.~Troessaert, \emph{{Finite BMS transformations}}, \href{https://doi.org/10.1007/JHEP03(2016)167}{\emph{JHEP} {\bfseries 03} (2016) 167} [\href{https://arxiv.org/abs/1601.04090}{{\ttfamily 1601.04090}}].

\bibitem{Donnay:2021wrk}
L.~Donnay and R.~Ruzziconi, \emph{{BMS flux algebra in celestial holography}}, \href{https://doi.org/10.1007/JHEP11(2021)040}{\emph{JHEP} {\bfseries 11} (2021) 040} [\href{https://arxiv.org/abs/2108.11969}{{\ttfamily 2108.11969}}].

\bibitem{Bondi:1962px}
H.~Bondi, M.G.J.~van~der Burg and A.W.K.~Metzner, \emph{{Gravitational waves in general relativity. 7. Waves from axisymmetric isolated systems}}, \href{https://doi.org/10.1098/rspa.1962.0161}{\emph{Proc. Roy. Soc. Lond. A} {\bfseries 269} (1962) 21}.

\bibitem{Newman:1966ub}
E.T.~Newman and R.~Penrose, \emph{{Note on the Bondi-Metzner-Sachs group}}, \href{https://doi.org/10.1063/1.1931221}{\emph{J. Math. Phys.} {\bfseries 7} (1966) 863}.

\bibitem{Madler:2016xju}
T.~M\"adler and J.~Winicour, \emph{{Bondi-Sachs Formalism}}, \href{https://doi.org/10.4249/scholarpedia.33528}{\emph{Scholarpedia} {\bfseries 11} (2016) 33528} [\href{https://arxiv.org/abs/1609.01731}{{\ttfamily 1609.01731}}].

\bibitem{SJFletcher:2003}
S.J.~Fletcher and A.W.C.~Lun, \emph{The kerr spacetime in generalized bondi–sachs coordinates}, \href{https://doi.org/10.1088/0264-9381/20/19/302}{\emph{Classical and Quantum Gravity} {\bfseries 20} (2003) 4153}.

\bibitem{Kapec:2017gsg}
D.~Kapec and P.~Mitra, \emph{{A $d$-Dimensional Stress Tensor for Mink$_{d+2}$ Gravity}}, \href{https://doi.org/10.1007/JHEP05(2018)186}{\emph{JHEP} {\bfseries 05} (2018) 186} [\href{https://arxiv.org/abs/1711.04371}{{\ttfamily 1711.04371}}].

\bibitem{He:2024vlp}
T.~He, A.-M.~Raclariu and K.M.~Zurek, \emph{{An infrared on-shell action and its implications for soft charge fluctuations in asymptotically flat spacetimes}}, \href{https://doi.org/10.1088/1751-8121/adc4a2}{\emph{J. Phys. A} {\bfseries 58} (2025) 165402} [\href{https://arxiv.org/abs/2408.01485}{{\ttfamily 2408.01485}}].

\bibitem{Poisson_2004}
E.~Poisson, \emph{A Relativist’s Toolkit: The Mathematics of Black-Hole Mechanics}, Cambridge University Press (2004).

\end{thebibliography}\endgroup
\end{document}